\documentclass{pasa}%

\title[The disc origin of the Milky Way bulge]{The disc origin of the Milky Way bulge}
\author[P. Di Matteo]{P. Di Matteo$^1$, \\
\affil{$^1$GEPI, Observatoire de Paris, PSL Research University, CNRS,
Univ Paris Diderot, Sorbonne Paris Cit\'e, Place Jules Janssen, 92195
Meudon, France}}%
\jid{PASA}
\doi{10.1017/pas.\the\year.xxx}
\jyear{\the\year}

\usepackage[authoryear]{natbib}
\bibpunct{(}{)}{;}{a}{}{,}
\usepackage{aas_macros}
\usepackage{amstext}
\usepackage{hyperref} 
\hypersetup{colorlinks,citecolor=blue,linkcolor=blue,urlcolor=blue}

\begin{document}%
\begin{abstract}

The Galactic bulge, that is the prominent out-of-plane over-density present in the inner few kiloparsecs of the Galaxy, is a complex structure, as the morphology, kinematics, chemistry and ages of its stars indicate. To understand the nature of its main components -- those at [Fe/H]$\gtrsim -1$~dex -- it is necessary to make an inventory of the stellar populations of the Galactic disc(s), and of their borders : the chemistry of the disc at the solar vicinity, well known from detailed studies of stars over many years, is not representative of the whole disc. This finding, together with the recent revisions of the mass and sizes of the thin and thick discs, constitutes a major step in understanding the bulge complexity. N-body models of a boxy/peanut-shaped bulge formed from a thin disc through the intermediary of a bar have been successful in interpreting a number of global properties of the Galactic bulge, but they fail in reproducing the detailed chemo-kinematic relations satisfied by its components and their morphology. It is only by adding the thick disc to the picture that we can understand the nature of the Galactic bulge.

\end{abstract}
\maketitle%
%

\section{THE GALACTIC BULGE}\label{bulge}

The bar-like shape and  boxy/peanut structure of the Galactic bulge, found also in a large number of external galaxies \citep{lutticke00, kormendy04}, have been known for long time \citep{okuda77, maihara78, weiland94, dwek95, binney97, babusiaux05, lopez05, rattenbury07, cao13}. This structure constitutes  the innermost thick part 
of a longer and flatter bar, extending up to about 4--5~kpc from the Galaxy centre \citep{benjamin05, cabrera07, wegg15}. Its orientation is between 20 and 30 degrees with respect to the Sun-Galaxy centre direction \citep{bissantz02, shen10, wegg13}, and at vertical distances from the plane of about 400 pc, a prominent X-shape becomes visible in the data \citep{mcwilliam10, nataf10, wegg13, gonzalez15clump}, indicative of a stellar bar evolving secularly after buckling \citep{combes81, athanassoula05, debattista06, martinezvalpuesta06, ness12, dimatteo14}.\\
Studies of the bulge kinematics also support its bar-like nature and thus ultimately its link to the disc. The bulge shows indeed cylindrical rotation \citep{howard08, howard09, kunder12}, as predicted by N-body models where the bulge is formed via the buckling of a pre-existing thin stellar bar formed in a cold stellar disc \citep{shen10, saha12, ness13kin, dimatteo15}. The vertex deviation observed in the stellar velocities \citep{zhao94, soto07, babusiaux10}, especially for its metal-rich population \citep{babusiaux10}, is also an indication of its asymmetric structure. \\

Bulge stars span a large range in [Fe/H] and [$\alpha$/Fe] values  \citep[with $\text{-3~dex} \le \text{[Fe/H]} \le \text{0.6~dex and 0~dex} \le \text{[$\alpha$/Fe]} \le \text{0.4~dex}$, see ][]{rich88,mcwilliam04, zoccali03, ness13spop, gonzalez15}, 
and several recent works show that their chemical patterns strikingly resemble those of stars at the solar vicinity \citep{melendez08, alves10, bensby10, ryde10, gonzalez11alpha, gonzalez15, ryde15, ness15}, once the outer disc sequence -- present at the solar vicinity and not in the inner disc (see next Section) -- is removed from the comparison. It is important here to note that previous works \citep{mcwilliam94, rich05, zoccali06, lecureur07, hill11, johnson11} have pointed out a difference between the chemistry of the bulge and that of the thick disc at the solar vicinity, with bulge stars being more $\alpha-$enhanced than thick disc stars of the same metallicity. 
This argument naturally led to exclude that the $\alpha-$enhanced population of the bulge could be the result of the secular evolution of the disc, via bar formation \citep[see for example][]{zoccali06}, as suggested also by chemical evolution models \citep{matteucci90, matteucci99, ballero07}.  However, in more recent studies, these differences appear much less dramatic than previous findings. Quoting \citet{johnson11}, for example: ``While [..] the [Si/Fe], [Ca/Fe], and [$\alpha$/Fe] ratios are generally enhanced by $\sim0.2$~dex in bulge giants compared to thick disk giants (and dwarfs), this enhancement is not extraordinarily
different than the combined measurement uncertainty and choice of abundance
normalization scale [...]. Interestingly, both the thick disk and bulge appear to
show similar declines in the [$\alpha$/Fe] ratio at [Fe/H]$\approx-0.3$." Indeed the current data show that the offset between the chemistry of  bulge stars and that of discs stars at the solar vicinity \citep[see][]{bensby10, gonzalez11alpha, adibekyan12, johnson14, ryde15}  is comparable to the existing systematics between spectroscopic studies of stars in the solar neighborhood \citep[compare, for example,][]{alves10, gonzalez11alpha, adibekyan12, bensby14} or between the differences among different bulge studies \citep{gonzalez11alpha, bensby10, johnson12, johnson14, vanderswaelmen16}. 

\begin{figure*}
\begin{center}
\includegraphics[clip=true, trim = 50mm 5mm 50mm 5mm, angle=270,width=2.\columnwidth]{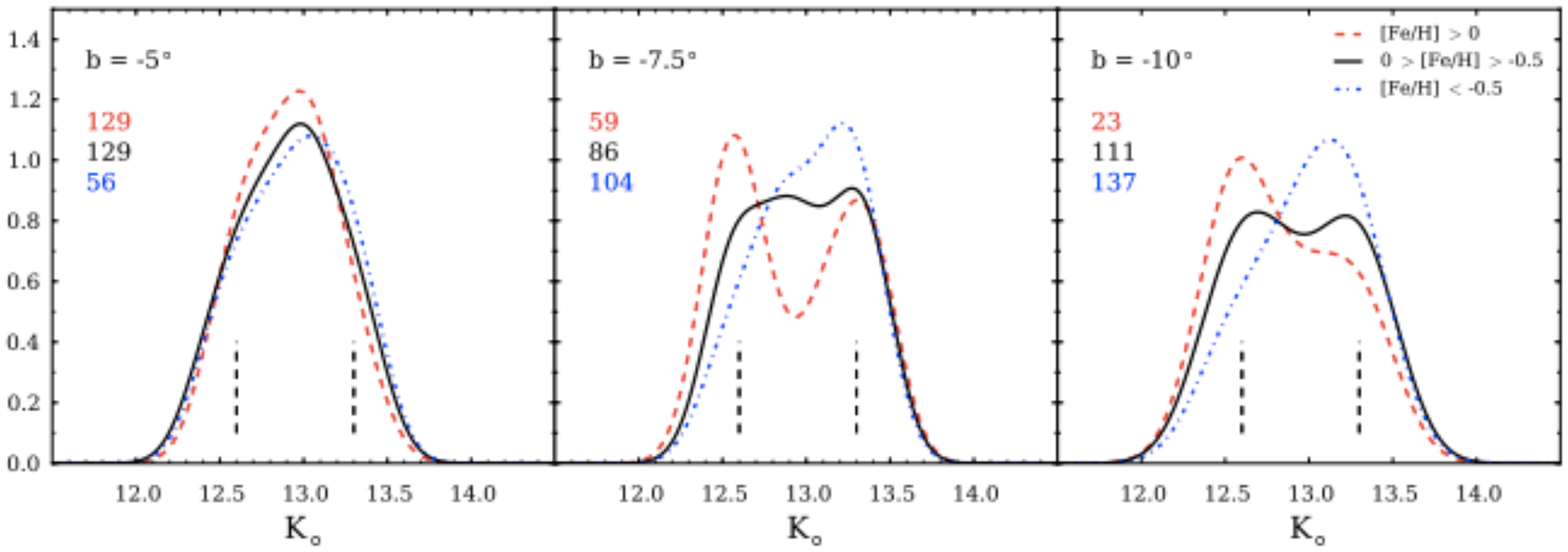}
\caption{From \citet{ness13spop}: Magnitude distribution of red clump stars along the bulge minor axis, at latitudes $\rm b=-5^\circ, -7.5^\circ \rm{and} b=-10^\circ$. The clump stars are splitter into three different metallicity bins : $\text{[Fe/H]} > 0$ (red curve), $-0.5~\text{dex}< \text{[Fe/H]}< 0$ and $\text{[Fe/H]} < -0.5~\text{dex}$. }\label{ness13fig}
\end{center}
\end{figure*}

Evidence about the presence of multiple components, or populations, in the bulge comes from the investigation of its metallicity distribution function (hereafter MDF), at different latitudes and longitudes \citep{babusiaux10, hill11, uttenthaler12, ness13spop, rojas14, gonzalez15}. In the largest spectroscopic study of the bulge conducted so far, the ARGOS survey \citep{freeman13}, \citet{ness13spop} decomposed the global MDFs found at latitudes $b=-5^\circ, -7.5^\circ, -10^\circ$ (integrated over the whole range of longitudes $-15^\circ\le l \le 15^\circ$) into five gaussian components, called A, B, C, D and E, whose metallicities  peak at about  [Fe/H]=0.1~dex, -0.3~dex, -0.7~dex, -1.2~dex and -1.7~dex, respectively.  Of those five components, D and E are negligible in the fields considered, representing 6\% at most of the whole sample, while the proportion of populations A, B and C has been shown to vary with the height from the plane: A  represents about 40\% of the sample at $b=-5^\circ$, and decreases to 16\% and 12\%, respectively, at $b=-7.5^\circ$ and $b=-10^\circ$; the  C component, on the opposite, increases with the height from the plane, passing from 20\% at $b=-5^\circ$ to 30\% at $b=-7.5^\circ$ and to roughly 40\% at $b=-10^\circ$; B shows the smallest variations, and its fractional contribution to the bulge is significant at all  latitudes considered in this study (40\% at $b=-5^\circ$, 50\% at $b=-7.5^\circ$, 44\% at  $b=-10^\circ$).
The different proportion of these components with height above the plane naturally explains the existence of a vertical metallicity gradient in the bulge, as reported by \citet{ness13spop} and even before in the studies of \citet{zoccali08, gonzalez11met, gonzalez13}. Interestingly, it has been shown that the structural and kinematic properties change from one component to the other \citep{ness13kin,ness13spop}. The most metal rich components A and B, characterized by [Fe/H]$\ge -0.5$~dex, show a split in the distribution of K-magnitudes of their red clump stars. This distribution is particularly suitable to understand the underlying morphology. The density of a peanut-shaped bulge indeed
shows some minimum in the centre, on its minor axis, minimum which is more accentuated at higher latitudes. On the line of sight, this produces a bimodal distribution, or a split, in the apparent K-magnitude of the red clump. In particular, it appears that in the Milky Way bulge the deepest minima between peaks in the K-magnitude distribution are found in the most metal-rich population \citep[i.e. population A, see Fig.~\ref{ness13fig} reproduced from][]{ness13spop}. At the latitudes covered by ARGOS, component C does not show any split. Interestingly, however, the maximum in the K-magnitude distribution of its stars is not located at K$=12.9$, as would be expected from a population which reaches its maximum density in the Galaxy centre, but rather it seems to peak  where the far lobe of the peanut of populations A and B is found  \citep[again, see Fig.~\ref{ness13fig} here reproduced from][]{ness13spop} .  I will comment more on this point in Sect.~\ref{models}.\\
At any given latitude, populations A, B and C also show different kinematics \citep{ness13kin}. In particular, while the rotation curve, as deduced from radial velocities, is very similar in the three populations at all latitudes\footnote{Note however that \citet{ness13kin} report a slightly higher (20\%) value of the rotational velocity for component B, when compared to A and C.}, showing that the cylindrical rotation observed globally is also reproduced by each component individually, the velocity dispersions change with latitude and depend on the population examined. Component  B has velocity dispersion profiles similar to A (but with higher absolute values), while component C shows dispersions constant both with longitude and latitude \citep[see Fig.~6 in][]{ness13kin}. As previously said, other studies have shown that the bulge MDF can be decomposed in multiple components. Depending on the number of components found \citep[two in][and five in \citet{ness13spop}]{babusiaux10, hill11, bensby11, uttenthaler12, rojas14, gonzalez15}, the fraction of stars attributed to the metal-rich and metal-poor component(s) change, and accordingly their structural and kinematic properties. \\
Finally, few words on the ages of stars in the bulge. Isochrone fitting  of main-sequence turnoff stars in the bulge have lead to the general conclusion that the majority of the stars in the bulge are old, typically $\gtrsim 10$~Gyr \citep{terndrup88, holtzman93, ortolani95, feltzing00, zoccali03, clarkson08, brown10, valenti13}.
Such old ages have been interpreted as evidence that the bulge is a classical spheroid, that is a structure formed by mergers in the early phases of the Milky Way evolution \citep{ortolani95, zoccali03, brown10}. However, some findings are questioning this picture.  The dating of microlensed dwarf stars in the bulge at latitudes $-5^\circ \le b \le -2^\circ$  has shown, indeed, that the most metal-rich stars  ([Fe/H]$> -0.4$~dex) show a range in stellar ages from 3 to 12 Gyr \citep[][, see also Fig.~1 in \citet{ness14}]{bensby13}. Young stars are not expected to be present  in a classical spheroid, formed at early times, but are a natural outcome of a bulge formed secularly, through the bar \citep{ness14}. Moreover, very recently  the VVV survey \citep{minniti10} has reported the presence of numerous classical Cepheids (ages $\le 100$~Myr) close to the Galaxy mid-plane  (latitudes $-1.7^\circ \le b \le 2^\circ$) and distributed along the whole bulge longitudinal extent \citep{dekany15}. The recent findings about the ages of stars in the bulge strengthen the global picture traced so far about the composite nature of this structure. \\Overall the bulge shows a very well defined vertical structuring:  the coldest, most metal-rich and youngest components are preferentially found at low latitudes and the weight of the dynamically hottest and most metal-poor components ($-1\le$[Fe/H]$\le -0.5$~dex) increases with height from the plane. A population with intermediate metallicity ($-0.5\le$~[Fe/H]~$\le 0$~dex, population B in \citet{ness13spop}) is however found at all latitudes in similar proportions. In the following of this review, I will discuss how it is possible to explain the trends observed for the main populations of the Galactic bulge -- those with [Fe/H]$\gtrsim -1$~dex-- in a scenario where these populations originate from the thin and thick discs of the Milky Way. I will not discuss further about stars with [Fe/H]$\lesssim -1$~dex, whose origin is very likely linked to the Galactic halo and/or the very early phases of Galactic formation \citep{diemand05, tumlinson10, garcia13, ness13spop, howes14, howes15}.

\section{THE GALACTIC DISC(S)}
\label{stellpop}

\begin{figure*}
\begin{center}
\includegraphics[clip=true, trim = 5mm 5mm 5mm 5mm, angle=270,width=1.2\columnwidth]{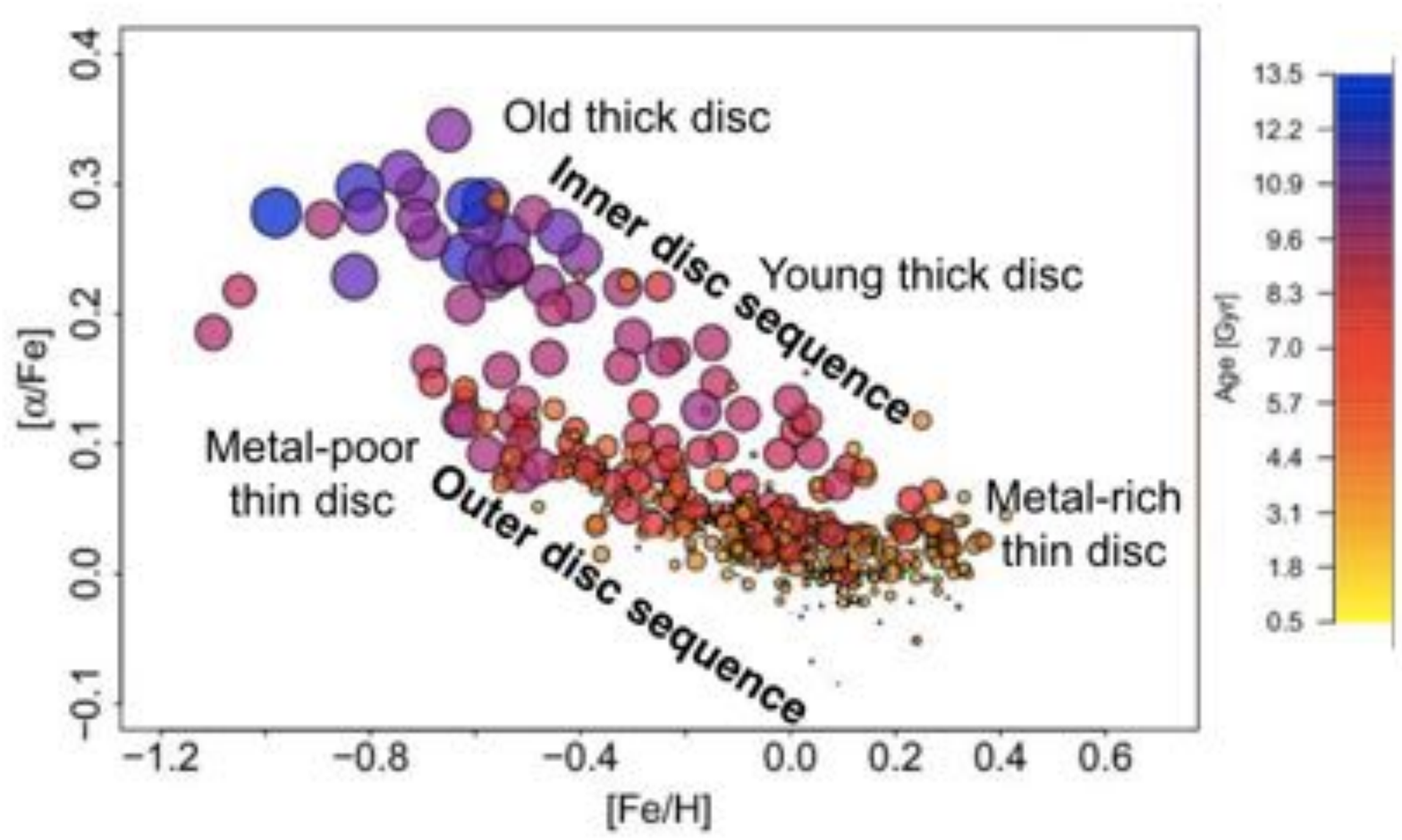}
\caption{[$\alpha$/Fe] versus [Fe/H] for the stars at solar vicinity  in the sample of \citet{adibekyan12} for which a robust age could be derived by \citet{haywood13}. The colors and the sizes of the symbols code the age of the stars. In this plot, the thick disc appears as  the sequence which extends from high [$\alpha$/Fe] and low metallicities until it joins the metal-rich thin disc, at solar [$\alpha$/Fe] and super-solar metallicities. In the text, this sequence is also called the ``inner disc sequence". The metal-poor thin disc extends from [Fe/H]$\sim$ -0.7~dex and super-solar [$\alpha$/Fe] to higher metallicities ([Fe/H]$\sim -0.3$~dex), but with the caveat that especially the upper  limit in metallicity of this sequence is very uncertain. This sequence represents that of the ``outer thin disc''. See \citet{haywood13} and discussion in the text for the reasons behind the use of this nomenclature.}\label{discscheme}
\end{center}
\end{figure*}

To understand the nature and origin of the Milky Way bulge it is necessary to locate it in its environment: the Galaxy, and in particular its stellar disc(s). 
To present our most recent view of the Galactic discs, I have chosen to maintain the classical nomenclature adopted in the literature. 

The \emph{thick disc} belongs to a sequence which extends from high [$\alpha$/Fe] and low metallicities until it joins the \emph{metal-rich thin disc}, at solar [$\alpha$/Fe] and super-solar metallicities. Because of the reasons recalled in the following of this section, I will refer to this sequence -- running from the metal-poor thick disc to the metal-rich thin disc -- also as the ``inner disc sequence''.  It is useful to remind at this point that this sequence shows a continuity in chemical patterns, with stellar ages decreasing with increasing metallicities  from about 13~Gyr -- for the oldest thick disc -- to ages of about 9--10~Gyr -- for the young thick disc -- to ages younger than 8~Gyr for the metal-rich thin disc \citep[see Fig.~\ref{discscheme} and][]{haywood13}. However, this continuity in chemical patterns hides a halt in the star formation history of the inner disc, which took place at the transition from the thick to the thin disc formation \citep{haywood16}. In the following, I use also the nomenclature ``inner thin disc" : this is essentially equivalent to the metal-rich thin disc, since -- as we will see in Sect.~\ref{thin} -- the thin disc in the inner regions reduces to the  metal-rich thin disc sequence only. \\
The \emph{metal-poor thin disc} extends from [Fe/H]$\sim-0.7$~dex and super-solar [$\alpha$/Fe] to higher metallicities ([Fe/H]$\sim -0.3$~dex), but with the caveat that especially the upper  limit in metallicity of this sequence is very uncertain. I will refer to this sequence also as that of the ``outer thin disc''. \\
The Sun is at the interface of the inner and outer thin discs, in a $\sim$3~kpc wide region that shares chemical properties of both the inner and outer discs. \\
The ``inner/outer disc" nomenclature used in this paper  has been already adopted in \citet{haywood13}, where we emphasized that ``there is perhaps more continuity between the thin and thick discs than between the inner thin disc and outer thin disc. The inner thin disc and the thick disc seem to be more like the same structure [...] while the outer disc appears more like a separate component" (see also Fig.~\ref{discscheme}). In the following, I will review the main reasons behind this vision, and explain why the most recent discoveries about the stellar populations of the Milky Way discs and their mutual links are of fundamental importance also for our understanding of the bulge, and its main components.
 
 \subsection{The thick disc}

The \emph{thick disc}, the last component discovered in the Milky Way in the early 80s \citep{gilmore83} has been for longtime seen as a population formed in a short-lived ($\sim$1 Gyr) starburst episode produced at early epochs \citep{burkert92, chiappini97, chiappini99, mashonkina01, fuhrmann04, reid05, bernkopf06}, chemically separated from the stellar thin disc \citep{fuhrmann98}, and with a rotational lag with respect to thin disc stars between 50 and 100 km/s \citep{norris86, wyse86, freeman87, ojha94, chiba00, fuhrmann08, bensby14}. With a scale length estimated to be similar or greater then that of the thin disc \citep{ojha01, robin03, juric08, dejong10} on the basis of color-selection studies, and with  a thick-to-thin disc volume density normalization  at the solar vicinity of 10\% at most \citep{robin03, juric08}, it has been long considered as a minor contributor to the mass budget of the Galaxy, constituting about 20\% of the thin disc mass \citep{robin03}.  We will see in the following that the estimates of thick disc scale lengths and mass have been significantly revised. But before moving to discuss these new estimates, I will recall the main processes proposed for the formation of thick discs in galaxies. 

Many different scenarios have been indeed suggested to explain thick discs formation. N-body models have shown that accretion of satellites can heat a kinematically cold disc and form, in few Gyrs, thick discs with vertical sizes, rotational lags, stellar eccentricities and masses similar to those observed for the Milky Way for appropriate orbital parameters, mass ratios and gas fractions of the progenitor galaxies \citep{quinn93, walker96, haung97, sellwood98, velazquez99, bekki01, hanninen02, read08, villalobos08, villalobos09, purcell09, moster10, qu10, qu11, house11, bekki11, dimatteo11, moster12}. One of the main objection to this scenario is that it should lead to a flaring of the disc \citep{bournaud09}, while usually these structures are not observed to flare \citep{vanderkruit82, degrijs98, comeron11}. Note however that recently it has been shown  that it is possible to reconcile an apparent lack of flaring with thick disc formed via satellite heating \citep{minchev15}.\\ Mergers of gas rich subunits, in which discs form thick ab initio, are also a viable mechanism of formation \citep{brook04, brook07, richard10}.\\ Clumpy disc galaxies, where massive gas and stellar over densities scatter stars in a pre-existing thin disc, making the overall disc thicker, have also been shown to be able to form thick structures in about 1 Gyr \citep{bournaud09}, as suggested by \citet{noguchi96}. However, this scenario may have difficulties in explaining the chemo-kinematic relations satisfied by thick disc stars in the Milky Way \citep[see][]{inoue14}.  \\Finally secular evolution processes, like radial migration, have been suggested to be able to explain the characteristics of the thick disc at the solar vicinity, namely its structure, kinematic and chemical properties \citep{schonrich09a, loebman11}. However, the effectiveness of this process in heating the disc has been repeatedly questioned  \citep{minchev12, veraciro14, halle15}.

\begin{figure}
\begin{center}
\includegraphics[clip=true, trim = 0mm 0mm 0mm 0mm, angle=0,width=1.\columnwidth]{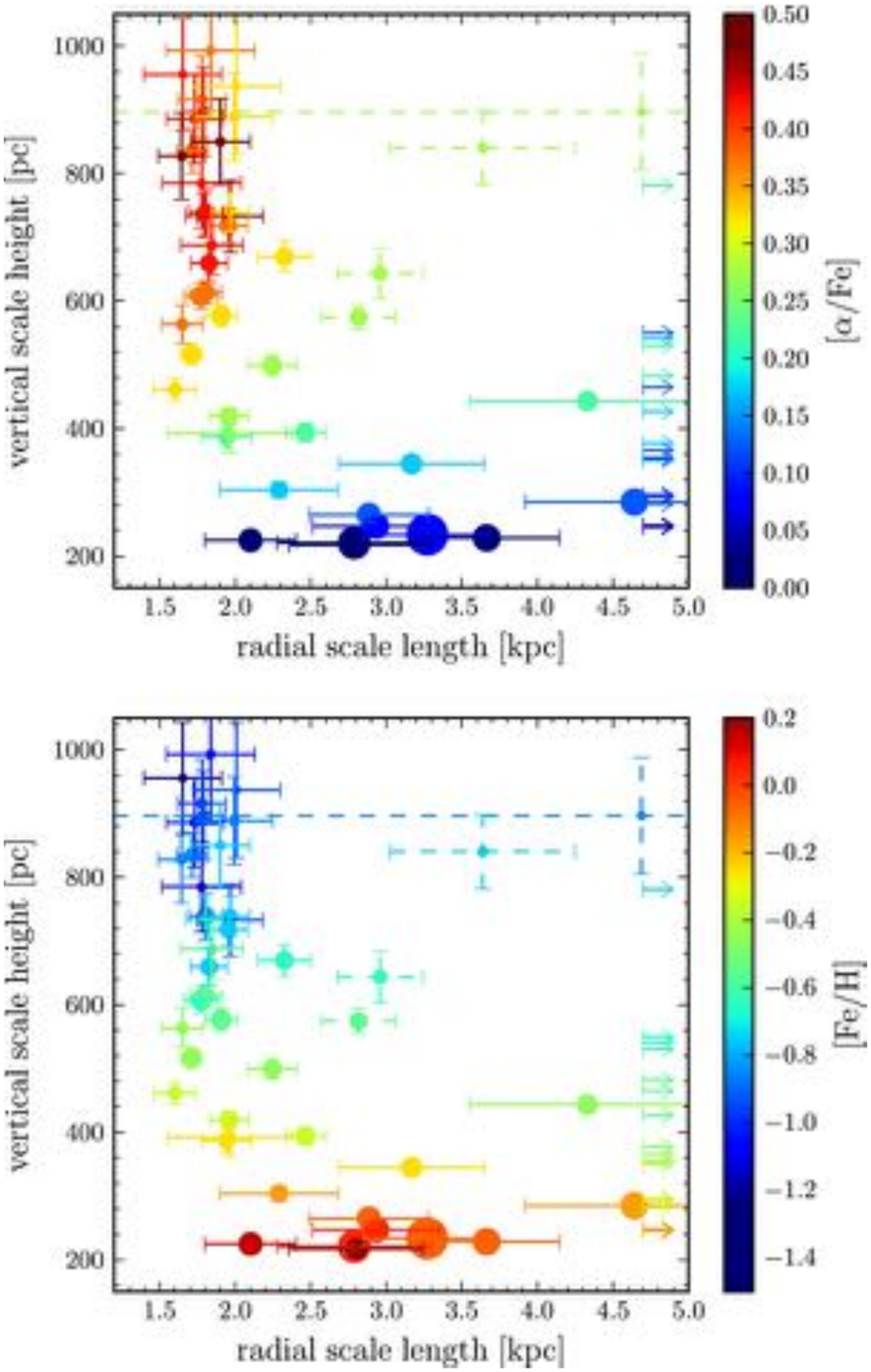}
\caption{From \citet{bovy12b}: vertical scale heights versus radial scale lengths for disc mono-abundance populations. In the top panel, points are color-coded by [$\alpha$/Fe], in the bottom panel by [Fe/H]. Alpha-abundant (i.e. [$\alpha$/Fe]$> \sim$ 0.2-0.3~dex in the \citet{bovy12b} scale), thick disc stars are characterized by short scale lengths, and have scale heights that diminish with decreasing [$\alpha$/Fe] and increasing [Fe/H].}\label{bovy12fig}
\end{center}
\end{figure}

These ``classical" picture and scenarios have been recently challenged by a number of findings. 
\begin{itemize}
\item Large scale spectroscopic surveys, like SEGUE and APOGEE, have shown the existence of a structural continuity between the thick and thin discs \citep{bovy12a, bovy12b, bovy15}, and inside the thick disc itself. The morphological-chemical structuring of the thick disc can be appreciated in Fig.~\ref{bovy12fig} reproduced from \citet{bovy12b}, where the vertical and radial scale lengths of disc stars grouped in bins of ([$\alpha$/Fe]-[Fe/H]) -- the so-called ``mono-abundance populations" in Bovy et al studies-- are plotted, with colors coding the [$\alpha$/Fe] and [Fe/H], respectively. Alpha-abundant (i.e. [$\alpha$/Fe]$> \sim$ 0.2-0.3~dex in the \citet{bovy12b} scale), thick disc stars are characterized by short scale lengths, and have scale heights that diminish with decreasing [$\alpha$/Fe] and increasing [Fe/H]\footnote{Note however that this trend of decreasing scale height with decreasing [$\alpha/Fe$] may not be valid for stars in the highest [$\alpha/Fe$] bins, see \citet{minchev14, guiglion15}}. Overall, the $\alpha-$enhanced thick disc has a scale length significantly shorter than that of the thin disc as a whole \citep[$\sim$2~kpc for the thick disc, and about 3.6~kpc for the thin; see][]{bovy12b, cheng12}. This scale length is about a factor of two smaller than previous estimates \citep{ojha01, juric08, dejong10}. That the thick disc had a shorter scale length than the thin disc was also found by means of high-resolution spectroscopic studies, which suggest a scale length for this population similar to that found by APOGEE and SEGUE \citep[see][]{bensby11}.  
\item  The analysis and dating of high spectroscopic data for stars at the solar vicinity have shown that the stars in the Galactic disc(s) satisfy a very well defined $[\alpha/\rm Fe]$-age relation and that the  $\alpha-$enhanced thick disc formed in a prolonged burst of star formation (lasting 3-4 Gyr, see \citet{haywood13}). Note that since those stars span a wide range of pericentre and apocentre distances, up to 1-2 kiloparsecs from the Galactic centre, this relation is not local but should apply to the whole thick disc \citep[see discussion in][]{haywood15}.
\item  The imprint left on the chemical abundances of long-lived stars at the solar vicinity coupled with a chemical evolution model, allowed to derive a  robust measurement of the Milky Way star formation history and to recognize that, in the Milky Way, the thick disc stellar population is as massive as its thin disc counterpart,  suggesting a fundamental role of this component in the genesis of our Galaxy \citep{snaith14, snaith15}.
\end{itemize}

That the thick disc is massive finds two independent confirmations.
\begin{enumerate}
\item In the most recent thick disc scale length measurements: as discussed in \citet{snaith14}, from the stellar surface densities derived from SEGUE \citep{bovy12b,bovy12c} and assuming that the thick and thin discs separate at [$\alpha$/Fe] = 0.25~dex \citep[on the SEGUE scale][]{bovy12b}, the thin disc contributes $\sim 21 \rm{M_\odot} \rm pc^{-2}$ to the local stellar surface density, and the thick disc $\sim 8 \rm M_\odot \rm pc^{-2}$.  Correcting for the scale length effect, the thick disc represents $\sim$47\% of the stellar mass within 10 kpc of the galactic centre, in accordance with the estimates derived from the star formation history and chemical modeling. \\
\item In the census of white dwarf stars at the solar vicinity : \citet{fuhrmann12} revised the fraction of local white dwarfs belonging  to the thin and thick discs, respectively, finding a larger fraction of thick disc white dwarfs than what suggested by previous estimates. This finding, coupled with the different scale lengths and heights of the two populations, lead \citet{fuhrmann12} to the conclusion that the thin disc is not dominant in the Milky Way, and that  very likely the two populations have equal masses.
\end{enumerate}

\emph{The most recent findings thus point to the presence of a massive thick disc, as massive as the thin counterpart, whose density at the solar vicinity is not representative of the global thick-to-thin mass ratio, and which is mainly concentrated in the inner disc.} Explaining the formation of such a massive and centrally concentrated component is of course challenging for any evolutionary model: it would require high mass fractions already in place at high redshift to sustain the prolonged star formation \citep{snaith14, snaith15, haywood15}, it is at odds with secular evolution models which only predict a marginal contribution of the thick disc to the overall mass budget of the Galaxy \citep{minchev12}, it requires mergers and accretion events to be fine tuned enough to generate (or preserve?) kinematics-chemistry(-age) relations as those observed today on local and kpc-scale data (see \citet{haywood13, bovy12a, bovy12b, bovy12c} and \citet{bird13, stinson13} for predictions from cosmological models).

 \subsection{The thin disc}\label{thin}

In the ``classical" picture, the \emph{thin disc} is the dominant and most massive stellar component of the Galaxy. Until few years ago, its kinematic and chemical properties were mostly known for stars at the solar vicinity. From kinematics studies, it has been inferred that the stars at the solar radius are significantly affected by the stellar Galactic bar : the Sun seems indeed to be located just outside the outer Lindblad resonance \citep[hereafter OLR, see][]{dehnen99, dehnen00, fux01, minchev07}, and several moving groups in the solar vicinity have been ascribed to the effect of bar resonances \citep{dehnen00, chakrabarty07, famaey07, famaey08, antoja09}. For stars confined at about a hundred parsecs from the Sun, where ages have been determined with sufficient accuracy, data show the existence of velocity dispersions - age relations for disc stars, with the older the stars the higher their dispersions \citep{wielen77, dehnen98, binney00, nordstrom04, seabroke07}, but note that some studies suggest a possible saturation for stars older than few Gyrs \citep{freeman91, edvardsson93, gomez97, quillen01, soubiran08}. These trends may be the signature of internal processes which kinematically heat stars over time \citep[e.g.][]{spitzer51, lacey84, carlberg85, kroupa02}. But they may also reflect a much more global trend observed in all disc stars, belonging to the thin as well as to the thick disc population. These two populations, examined as a whole, show a continuous decrease of velocity dispersions with age, from the oldest thick disc stars to the youngest thin disc objects \citep{haywood13}. However, it is not clear how representative this relation is of the initial formation of the disc(s) and how much internal processes (heating by spiral arms, bar formation and buckling, etc) and accretion events may have changed it over time, also because different evolutionary scenarios may be able to produce similar relations at the present epoch \citep{house11}. \\
From the chemical point of view, the thin disc shows signatures of a complex formation history. Stars at the solar vicinity have a metallicity distribution which peaks around [Fe/H]$\sim$0 \citep{haywood01}, with the tails of the distribution extending from [Fe/H]$\sim$-0.8~dex to super solar values. The large spread in metallicities of stars at the solar vicinity may be appreciated also looking at the age-metallicity relation: at any age, the dispersion is too significant to be explained simply by inhomogeneities in the interstellar medium at the time of formation of the different generations of stars \citep{haywood08, haywood13}. Some other processes have been thus invoked. Radial migration \citep{roskar08, brunetti11, minchev11, brook12, minchev13, miranda15} has been, over the last ten years, the most popular scenario to explain the metallicity and abundance distributions of stars at the solar vicinity, starting from the early suggestion by \citet{sellwood02}. N-body models have repeatedly found that radial migration \citep[both blurring and churning, following the terminology of][]{schonrich09b} is a global process, stimulated by the presence of one or multiple stellar asymmetries, and as such it should have naturally affected the whole disc. However, this seems at odds with a number of more recent findings \citep[see, for example,][and the following of this section]{haywood13, halle15} that question the strength of radial migration and the spatial extent over which it  is able to operate in our Galaxy. 
Many studies are indeed revealing that the thin disc appears radially structured, with little contamination between the inner thin disc ($R_{GC}\le 6-7$~kpc, with $R_{GC}$ being the in-plane distance from the Galactic centre) and the outer thin disc ($R_{GC}\ge10$~kpc). It is only in an intermediate region containing the solar radius ($6-7$~kpc$\le R_{GC}\le 10$~kpc) that the chemical patterns of the inner and outer disc coexist. This result is fundamental not only because it questions the borders of radial migration, but also for the comprehension of the bulge components. For this reason, I will devote the rest of the section in revising the evidence of this inner-outer disc dycothomy.  

\begin{figure}
\begin{center}
\includegraphics[clip=true, trim = 55mm 30mm 55mm 30mm, width=1.\columnwidth]{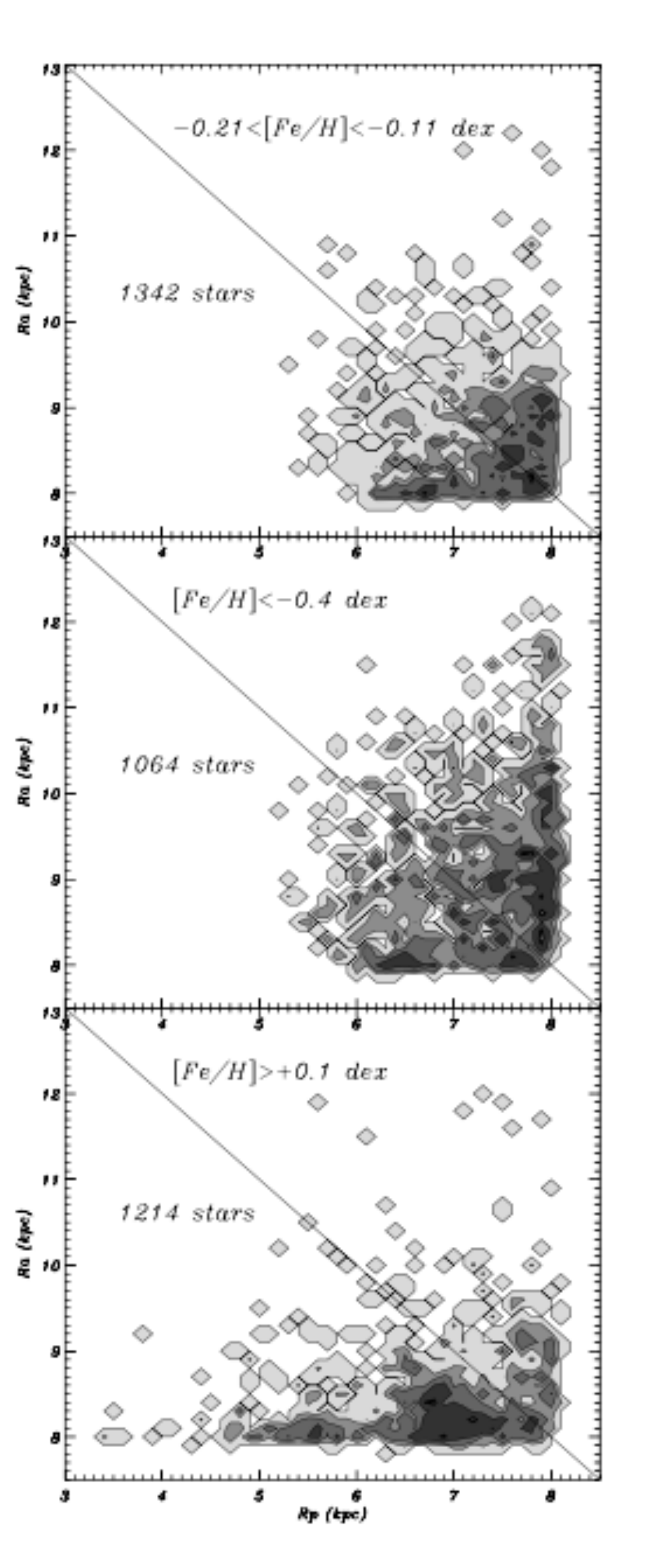}
\caption{From \citet{haywood08}: The different distribution of pericentres and apocentres  for stars at the solar vicinity kinematically selected as thin disc stars in the Geneva Copenhagen survey \citep{nordstrom04}, for different metallicity intervals. As evidenced by \citet{haywood08}, thin disc stars at the solar vicinity do not all share the same orbital properties: most of the metal-rich thin disc stars ([Fe/H]$\ge0.1$~dex)  tend to have pericentres inside the solar radius, up to few kpc from the Galactic centre, and apocentres at or just outside the solar position, while  metal-poor thin disc stars ([Fe/H]$\le -0.4$~dex) have orbits with pericentres above 6~kpc and with apocentres which extend much further out in the disc, up to $\sim$12~kpc from the Galaxy centre. }\label{haywood08fig}
\end{center}
\end{figure}

\begin{figure*}
\begin{center}
\includegraphics[clip=true, trim = 20mm 5mm 40mm 5mm, angle=270,width=2.\columnwidth]{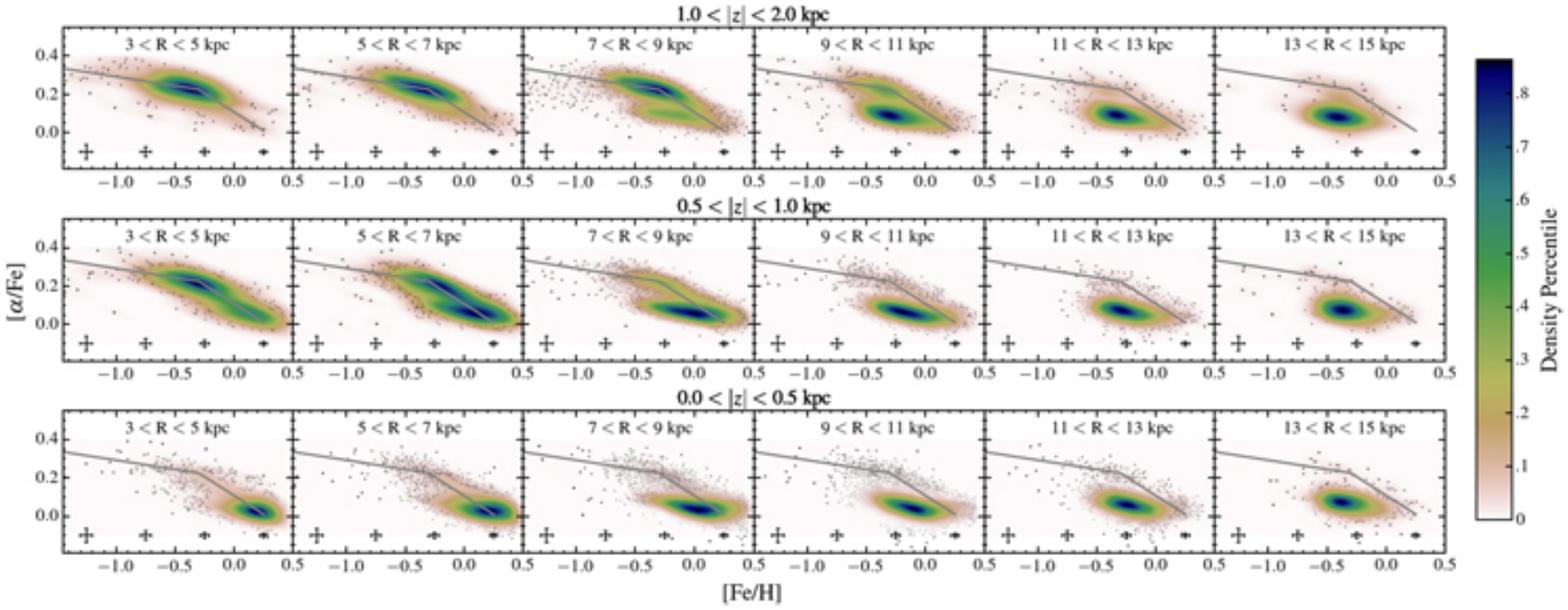}
\caption{From \citet{hayden15} : Distribution of stars in the [$\alpha$/Fe] versus [Fe/H] plane as a function of R and $|\text z|$ as revealed by APOGEE. The black line in each panel shows  the similarity of the shape of the $\alpha-$abundant sequence with R. Different rows correspond to stars at different heights above the Galactic plane, different columns to stars in different radial bins.}\label{hayden15fig}
\end{center}
\end{figure*}

A first indication comes from the work of \citet{haywood08}, who shows that thin disc stars at the solar vicinity do not all share the same orbital properties: most of the metal-rich thin disc stars ([Fe/H]$\ge0.1$~dex) found at the solar vicinity tend to have pericentres inside the solar radius, up to few kpc from the Galactic centre, and apocentres at or just outside the solar position, while  metal-poor thin disc stars ([Fe/H]$\le -0.4$~dex) have orbits which do not penetrate so much in the inner disc (their pericentres are usually above 6~kpc) and their apocentres extend much further out in the outer disc, up to 12 kpc from the Galaxy centre (see Fig.~\ref{haywood08fig}).

A second indication comes from the work of \citet{bensby11} : this study reveals that the (thin+thick discs) chemical patterns found at the solar radius are not representative of the whole disc. Only at the solar vicinity the two patterns -- the one going from [$\alpha$/Fe]$\sim$0.4~dex and [Fe/H]$\sim$-1~dex to solar values, and the one going from [$\alpha$/Fe]$\sim$0.2~dex and [Fe/H]$\sim$-0.7~dex (red and blue lines, respectively, in Fig.~2 of their work)-- coexist. In their inner disc sample, made of stars with distances from the galactic centre between 4 and 7~kpc, the second pattern is not present, while it is the only pattern that appears in the outer disc (distances between 9 and 13~kpc). In particular, it is striking to observe that metal-rich thin disc stars, present at the solar vicinity, are not found in the outer disc, at distances greater than $\sim9$~kpc from the galaxy centre. 
In \citet{haywood13}, we have re-investigated the question, calculating the ages of a high resolution spectroscopic sample of stars at the solar vicinity \citep[from][]{adibekyan12}. From this study, we concluded that the inner ($R_{GC}\le6-7$~kpc) and outer ($R_{GC}\ge10$~kpc) discs did not follow the same chemical evolution: at a given age, outer disc stars are more $\alpha-$enriched than co-eval inner disc stars, and have metallicities up to $\sim 0.4$~dex lower. 
Having detailed chemical abundances and ages, in \citet{snaith14, snaith15}, we reconstructed the star formation histories of the inner and outer discs, showing that the outer disc started to form about 10~Gyr ago, from an interstellar medium whose metallicity was diluted by 0.3--0.4~dex with respect to the metallicity of the interstellar medium present in the inner disc at the same time. In \citet{haywood13}, we proposed that the outer regions of the Milky Way possibly formed from a mixture of enriched material expelled from the thick disc, and of accreted, metal-poor gas.  Afterwards, the evolution of the outer disc remained essentially disconnected from that of the inner disc.
All these studies, except that of \citet{bensby11}, have been based on the analysis of stars at the solar vicinity. The definite confirmation that this picture is correct, and that the chemical patterns of the inner and outer discs are different and co-exist only in the solar region comes from APOGEE. The distribution in the [$\alpha$-Fe]-[Fe/H] plane of about 70,000 stars in the APOGEE survey, spanning a large range of distances from the Sun, is shown in \citet{hayden15} and here reproduced in Fig.~\ref{hayden15fig}. This figure perfectly summarizes the points made in this section :
\begin{itemize}
\item the $\alpha-$enhanced thick disc is centrally concentrated -- very few thick disc stars are indeed observed at distances greater than 10--11~kpc from the Galaxy centre;
\item in a 2-kpc wide region around the Sun, two chemical patterns are observed : one going from $\alpha-enhanced$, metal-poor thick disc stars to solar [$\alpha$/Fe] and super--solar  [Fe/H] values, the second pattern  starting at [$\alpha$/Fe]$\sim0.1$~dex and [Fe/H]$\sim -0.5$~dex with [Fe/H] increasing up to solar values. These two patterns co-exist only in a few kpc region around the Sun : in the outer disc the $\alpha-$enhanced pattern disappears, while in the inner disc, it is the low-$\alpha$ sequence which is not present.
\end{itemize}

There are at least two implications of these results which are of interest for our study:
\begin{itemize}
\item Radial migration by churning  \citep{sellwood02, schonrich09b, roskar08,  minchev13}, that is the change of guiding radii of stars caused by the presence of one or multiple asymmetries in the stellar disc, has not been efficient in mixing stars of the inner ($R_{GC}\le6-7$~kpc) and outer  Milky Way disc ($R_{GC}\ge9-10$~kpc). The difference of chemical patterns of these two regions of the Galaxy, with an overlap only in a few-kpc zone around the Sun, implies that no siginificant migration  of stars has occurred between these two regions. Whatever the formation scenario for the outer disc of the Milky Way is (gas/satellites accretion with a possible role of winds from the inner disc), and whatever the process that ignited star formation in those outer regions about 10~Gyr ago, some mechanism has prevented stars from migrating afterwards. In \citet{halle15}, we propose that the mechanism that inhibited migration is dynamical and is related to the position of the bar OLR, position that is currently estimated to be just inside the solar radius. By using N-body models, we show indeed that the bar OLR limits the exchange of angular momentum, separating the disc in two distinct parts with minimal exchange, except in the transition zone, which is delimited by the position of the OLR at the epoch of the formation of the bar, and at the final epoch. This also implies that stars migrated in the bar/bulge region  from the outer disc can originate in a region whose maximal extent is given by the final (i.e. current) position of the OLR \citep{dimatteo14}. Unless the Galaxy, over time, has experienced the formation  of several bars, and/or strong spiral waves, with different pattern speeds (in that case the scenario would become much more complex), the picture is simple : \emph{only stars inside the solar region, where the OLR should be currently located, can have migrated in the inner disc and contributed to the bulge populations.}
\item \emph{The chemical patterns of the inner disc are only part of the chemical patterns observed at the solar radius. When one compares the chemistry of bulge stars to those of disc stars, it is fundamental to take this result into account. } In particular, if the comparison is done with the chemical patterns observed at the solar vicinity, the low-$\alpha$ pattern of the metal-poor thin disc must not be taken into account in the comparison, simply because this pattern is typical of the outer disc and not of the region inside 6--7~kpc from the Galaxy centre. \emph{The chemical patterns of the bulge differ from those of the stellar disc at the solar vicinity as much as the latter differs from the chemical patterns of the inner disc}. The difference in chemical patterns of the bulge and of the thin disc at the solar radius arises because the solar radius is not representative of the chemistry of the inner thin disc, not because the metal-rich bulge is a different population from the inner thin disc. In this respect, the $\alpha-$enhancement of the bulge with respect to the thin disc found, for example,  by \citet{johnson11, gonzalez15} arises because  the whole thin disc is included in the comparison, while the metal-poor thin disc present at the solar vicinity should be removed. 
Once this is done, the filiation between the inner thin disc and the metal-rich bulge becomes evident. 
\end{itemize}
 
\section{THE BULGE/BAR/DISC(S) CONNECTION IN N-BODY SIMULATIONS}\label{models}

The observational evidences recalled in the previous sections suggest to investigate the nature of the bulge and its stellar components in view of the recent discoveries about the disc(s) populations (Sects.~\ref{bulge} and \ref{stellpop}). 
It is by quantifying and understanding the response of a thin+thick disc to bar formation and instabilities that we can solve the puzzle of the main (i.e. [Fe/H]$\ge-1$) populations of the Galactic bulge. 

In the following of this section, I will discuss three points, that are fundamental to me to understand why our current modeling of the Milky Way bulge is incomplete.  
\begin{enumerate}
\item N-body simulations which try to interpret the Milky Way bulge as the result of a bar instability in a pure thin disc galaxy fail in reproducing the chemo-kinematic relations, as found in the most recent surveys. They are successful in reproducing global trends, but not the trends satisfied by the individual bulge components. This  failure has been evidenced in \citet{dimatteo15}, and I will summarize here the main points made in that paper, after recalling the successes of the bulge/bar/thin disc scenario.
\item If the Milky Way contains any classical bulge, its mass must be small (not more than 10\%). Several independent N-body models have now reached this same conclusion, and this also means that the origin of the metal-poor populations which are significant in the bulge -- 60\% and more, depending on latitude, for stars with [Fe/H]$\le 0$~dex; up to 40\% for stars with [Fe/H]$\le-0.5$~dex only; according to \citet{ness13spop} -- cannot be ascribed to the presence of a classical bulge, which makes 10\% at most of the thin disc mass.
\item Neither a pure thin disc model nor the inclusion of a low-mass classical bulge can explain the properties of the metal-poor component present in the bulge. It is only by adding the thick disc to the picture that we can naturally  understand the vertical structuring of the bulge, the variation of metal-rich and metal-poor components with height, their relative weight, and the absence of a boxy/peanut-shaped structure for the metal-poor ($-1$~dex$\le$[Fe/H]$\le -0.5$~dex) stars at latitudes $b \ge -10^\circ$. I will show this, by presenting the first results of N-body simulations which include a massive thick disc, and that have high enough resolution to allow to study in detail the mapping of a thin+thick disc into a boxy bulge.
\end{enumerate}

\subsection{The bar/bulge/thin disc connection : successes and failures}

\begin{figure}
\begin{center}
\includegraphics[clip=true, trim = 15mm 40mm 5mm 40mm, width=1.\columnwidth, angle=0]{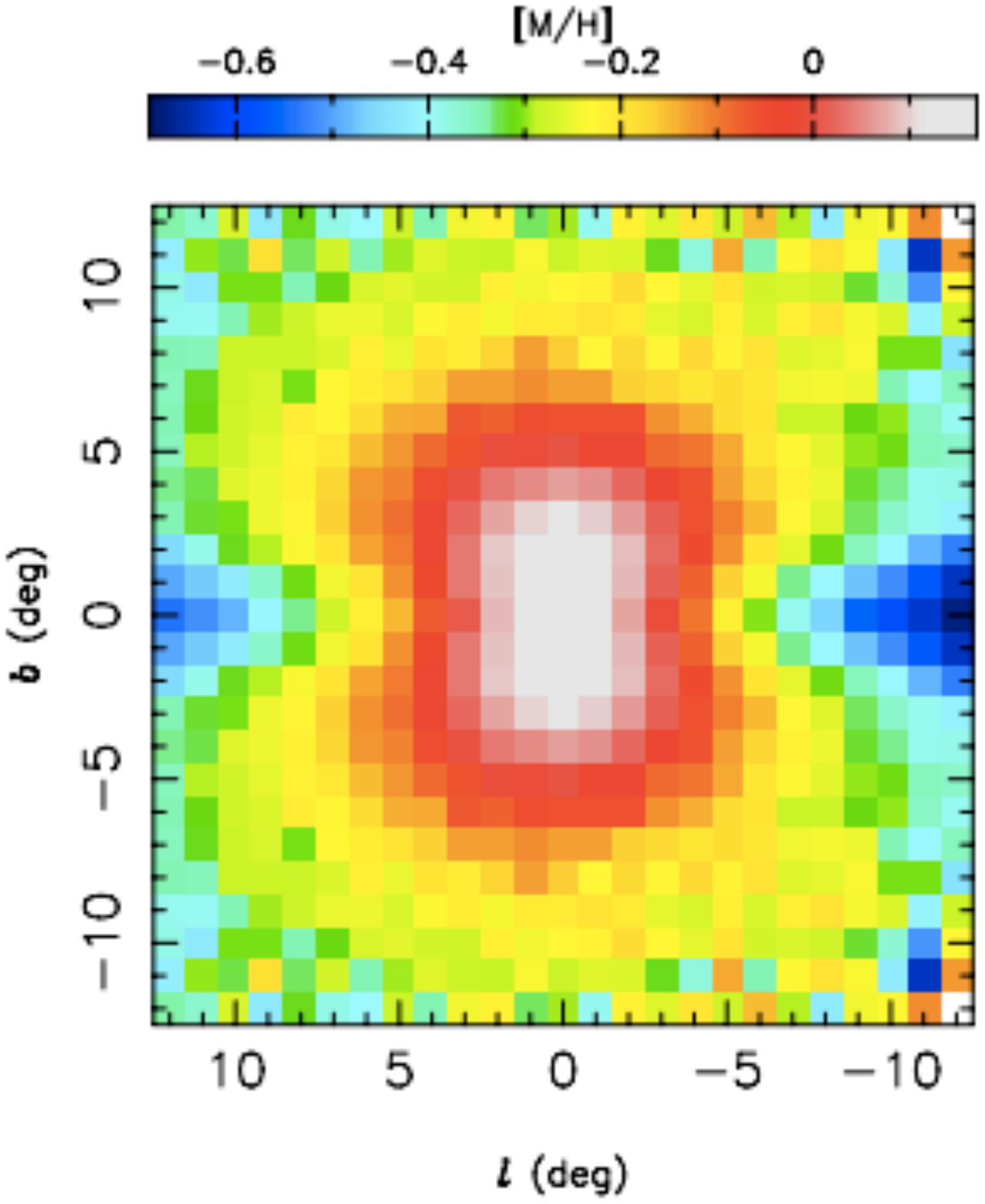}
\caption{From \citet{martinez13}: Bulge metallicity maps of a simulated thin disc galaxy in galactic coordinates ($l,b$). The modeled galaxy has initially a steep radial metallicity gradient in the disc, that is mapped into a vertical metallicity gradient in the bulge, when the bar buckles. The trends found are remarkably similar to observations \citep{gonzalez13}. However, even if successful in reproducing global bulge properties, a pure thin disc model for the Galactic bulge --as the one presented here -- fails in reproducing the detailed chemo-kinematic relations of its stars (see Fig.~\ref{dimatteo15fig2} and \citet{dimatteo15}).}\label{dimatteo15fig}
\end{center}
\end{figure}

\subsubsection*{Successes}
Because of the boxy/peanut-shaped nature of the Milky Way bulge and because of the marginal role that the thick disc was considered to play in the mass budget of the Galaxy until recently,  all N-body models so far -- except those of \citet{bekki11} -- have modeled the visible Galaxy as a pure thin disc, with the possible inclusion of a classical bulge.
These models, which I will refer to also as ``pure thin disc" models -- even in the case they contain a small ($\sim 10\%$) classical bulge, as in \citet{shen10, dimatteo15} --  have been successful in reproducing a number of global characteristics of the Milky Way bulge, as its structure and X-shape, its kinematics and metallicity trends.\\
\citet{martinez11} and \citet{gerhard12}, for example, showed how a pure thin disc model can be able to reproduce star counts observations in the Galactic plane, for the long bar, and in the inner  regions of the boxy bulge ($-10^\circ \le l \le 10^\circ$ and $b=\pm1^\circ$), where a sudden change in the structure of the bulge is observed at $|l|\sim 4^\circ$ \citep{nishiyama05, gonzalez11c}. By means of a N-body model of a stellar bar grown from a thin disc, \citet{ness12} were able to explain the split  in the distribution of red giant clump stars along the bar minor axis, as observed by ARGOS.

The pure thin disc model developed by \citet{shen10} has been successful in explaining the stellar kinematics of the bulge region from the BRAVA bulge survey \citep{howard08}, at latitudes $b=-4^\circ, -6^\circ$ and $-8^\circ$ and longitudes $-10^\circ \le b \le +10^\circ$. In particular, this N-body simulation reproduces the cylindrical rotation observed all over the analyzed fields and the corresponding trends in velocity dispersions. \citet{kunder12} confirmed this agreement,  by comparing \citet{shen10} model to a more extended sample of stars from BRAVA at similar latitudes and longitudes. Similar results were obtained by \citet{gardner14}, in their comparison with BRAVA, and by \citet{zoccali14}, who used the pure thin disc N-body
model presented in \citet{martinez11} and compared it with the bulge kinematics from the GIBS survey, extending the comparison between data and models also closer to the Galaxy mid-plane, at latitude $b=-2^\circ$. 
That a ``pure thin disc" model is able to reproduce the global kinematics of the bulge has been also shown by \citet{ness13kin}, in their comparison with the ARGOS kinematics at latitudes $b=-5^\circ, -7.5^\circ$ and $-10^\circ$  and by \citet{dimatteo15}, who compared their N-body model to ARGOS and BRAVA data.

Such models have been shown to be able to reproduce also the vertical metallicity gradient, similar to those  observed \citep{zoccali08, gonzalez11met, gonzalez13, ness13spop}. \citet{martinez13} indeed showed how an initial steep ($\sim -0.4$~dex/kpc) radial metallicity gradient in the thin disc can be mapped into a comparable vertical metallicity gradient in the bulge: the most loosely bound stars, in the disc outskirts, are indeed preferentially redistributed at large heights from the plane, while more bound stars tend to be found at low latitudes \citep[see also][]{dimatteo14}. This explain why, in a pure thin disc which has initially -- before bar formation --  a steep enough metallicity gradient, models predict also longitude-latitude metallicity maps remarkably similar to observations \citep[see Fig.~\ref{dimatteo15fig} and][]{martinez13}. I will comment in the following on the limitations of this scenario.\\
Finally, even if we are only starting to discover the spatial redistribution of stars with different ages in the Galactic bulge, the most recent findings \citep{bensby13, dekany15}  seem in agreement with models predictions about the presence of young stars close to the Galaxy mid-plane, as a consequence of the continuous star formation in the galactic thin disc \citep{ness14}. \\

\begin{figure}
\begin{center}
\includegraphics[clip=true, trim = 15mm 10mm 5mm 10mm, width=1.\columnwidth, angle=0]{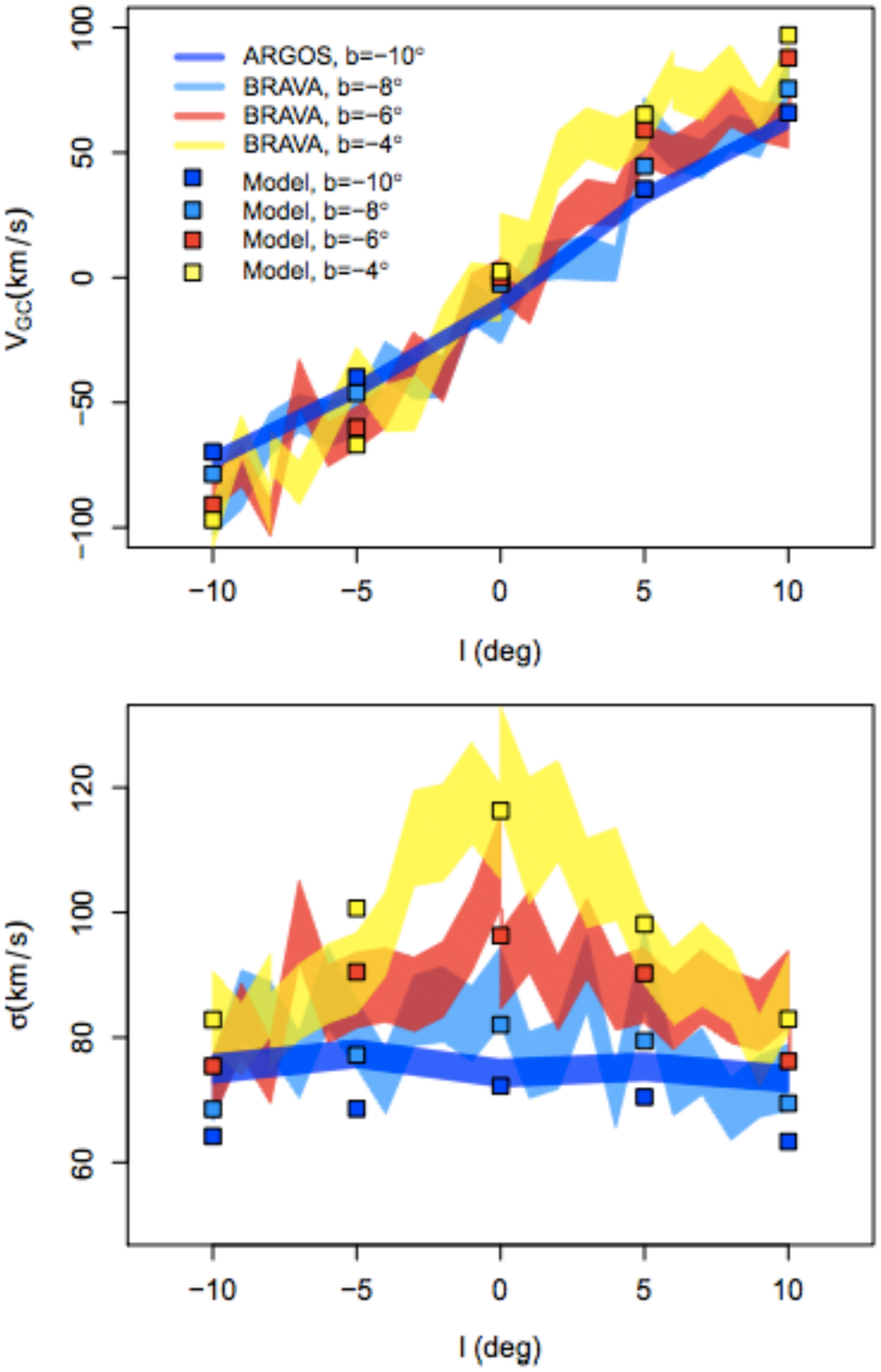}
\caption{From \citet{dimatteo15}:  Rotation curve (top panel) and velocity dispersions (bottom panel) of bulge stars in the pure thin disc  N-body model discussed in \citet{dimatteo15}.  Four different latitudes are shown for the modeled galaxy: $b = -4^\circ$ (yellow squares), $b = -6^\circ$ (red squares), $b = -8^\circ$ (pale blue squares), $b = -10^\circ$ (dark blue squares). For comparison, BRAVA fields at $b = -4^\circ$ (yellow, solid curve), $b = -6^\circ$ (red, dashed curve), and $b = -8^\circ$ (pale blue, dotted curve), and ARGOS fields at $b = -10^\circ$ (dark blue, dash-dotted curve) are also given. The thickness of the curves corresponds to the $\pm 1\sigma$ error in the observational data. See also \citet{shen10, kunder12, zoccali14} for similar global kinematic trends.  However, even if successful in reproducing global kinematic properties, a pure thin disc model -- as the one presented here -- for the Galactic bulge fails in reproducing the detailed chemo-kinematic relations of its stars (see Fig.~\ref{dimatteo15fig2} and \citet{dimatteo15}).}\label{dimatteo15fig1}
\end{center}
\end{figure}

\begin{figure*}
\begin{center}
\includegraphics[clip=true, trim = 30mm 70mm 25mm 75mm, width=1\columnwidth, angle=0]{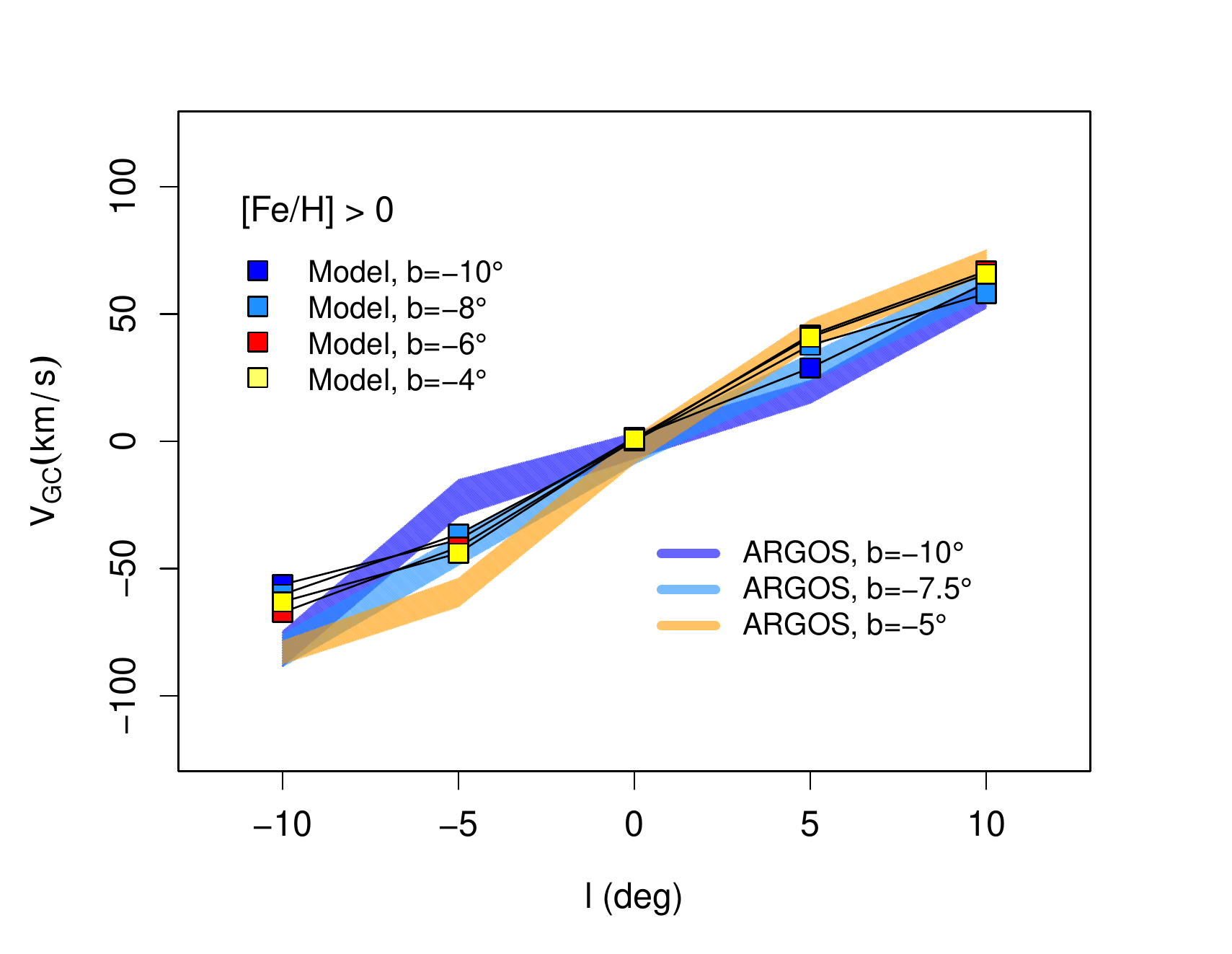}
\includegraphics[clip=true, trim = 30mm 70mm 25mm 75mm, width=1\columnwidth, angle=0]{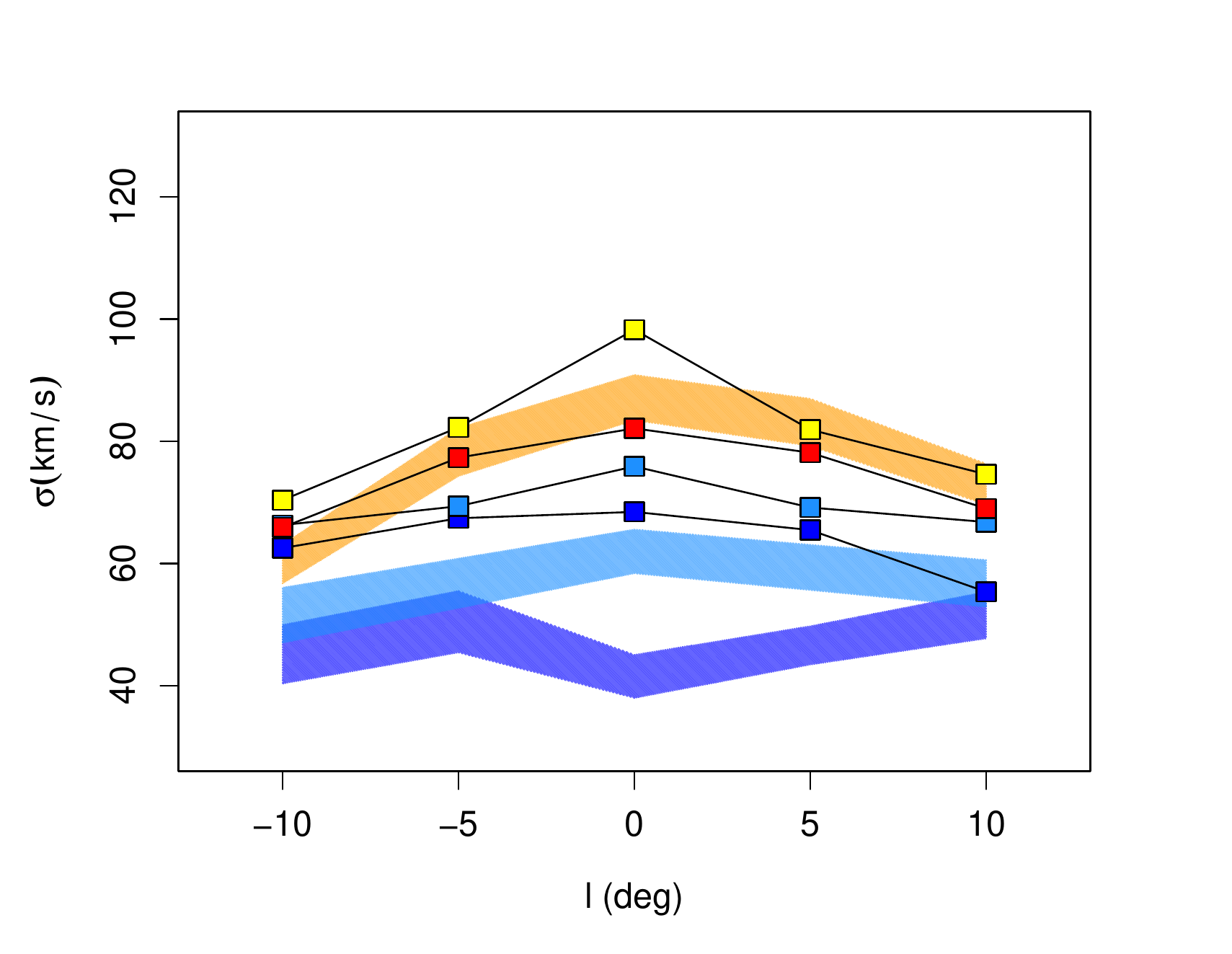}
\includegraphics[clip=true, trim = 30mm 70mm 25mm 75mm, width=1\columnwidth, angle=0]{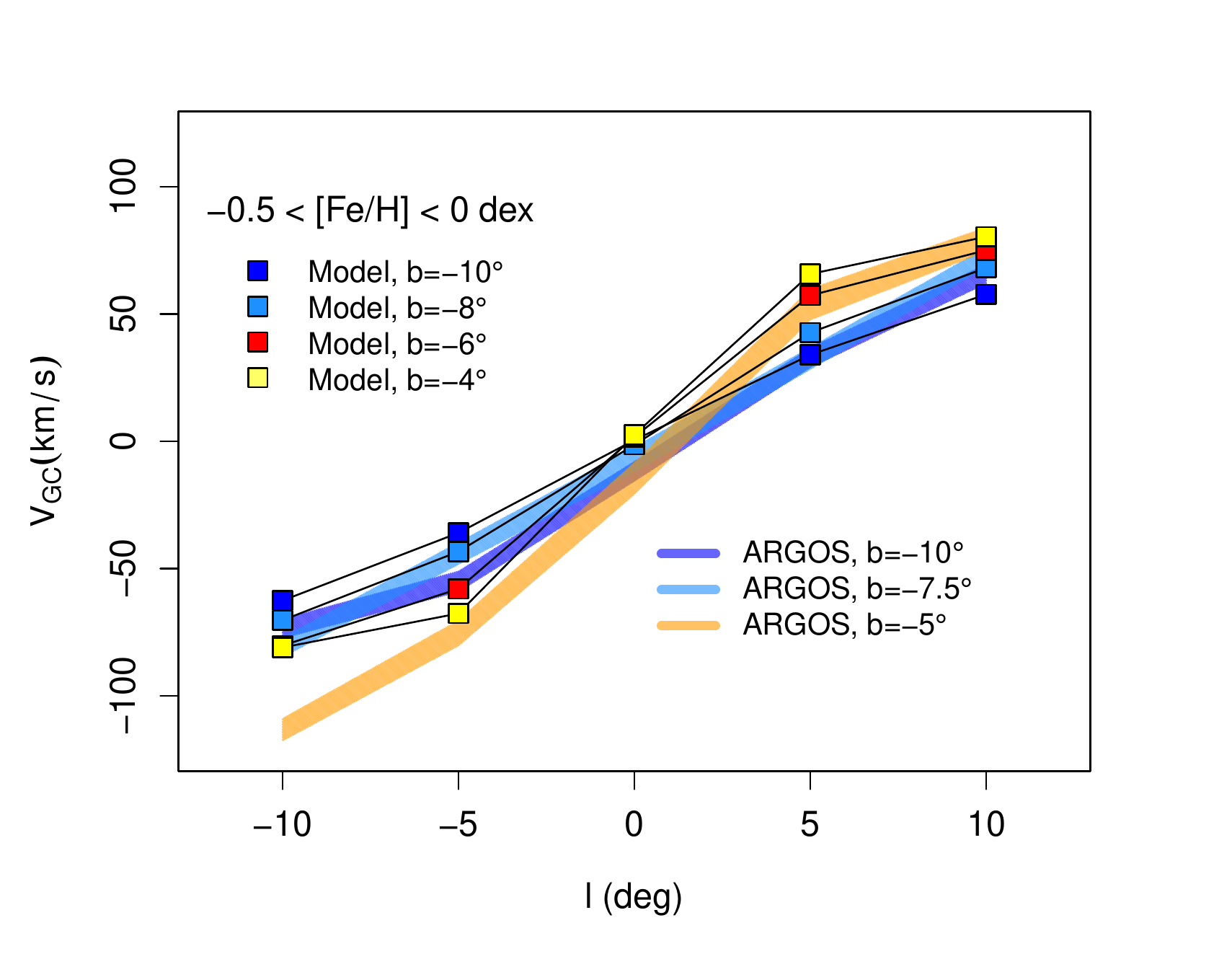}
\includegraphics[clip=true, trim = 30mm 70mm 25mm 75mm, width=1\columnwidth, angle=0]{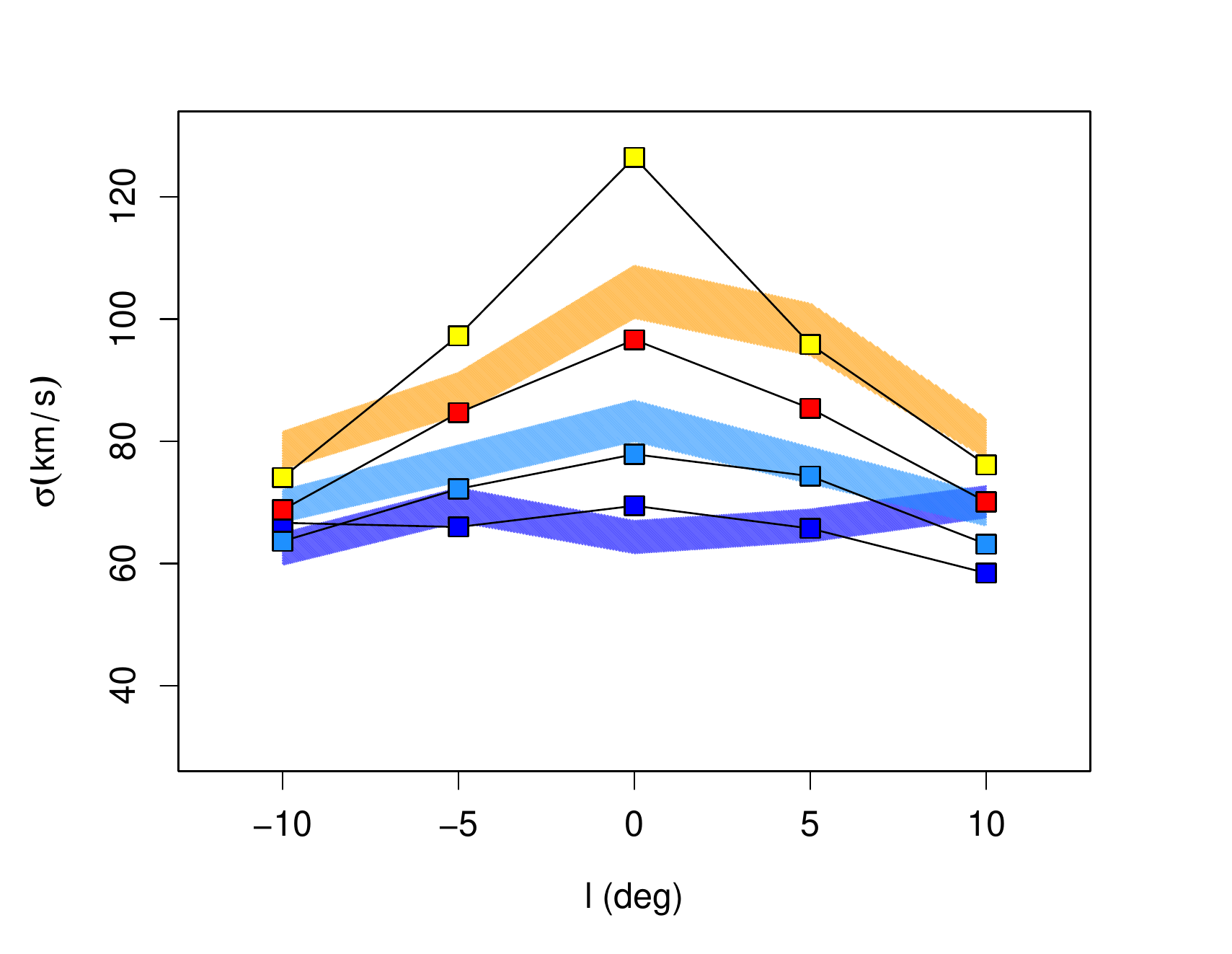}
\includegraphics[clip=true, trim = 30mm 70mm 25mm 75mm, width=1\columnwidth, angle=0]{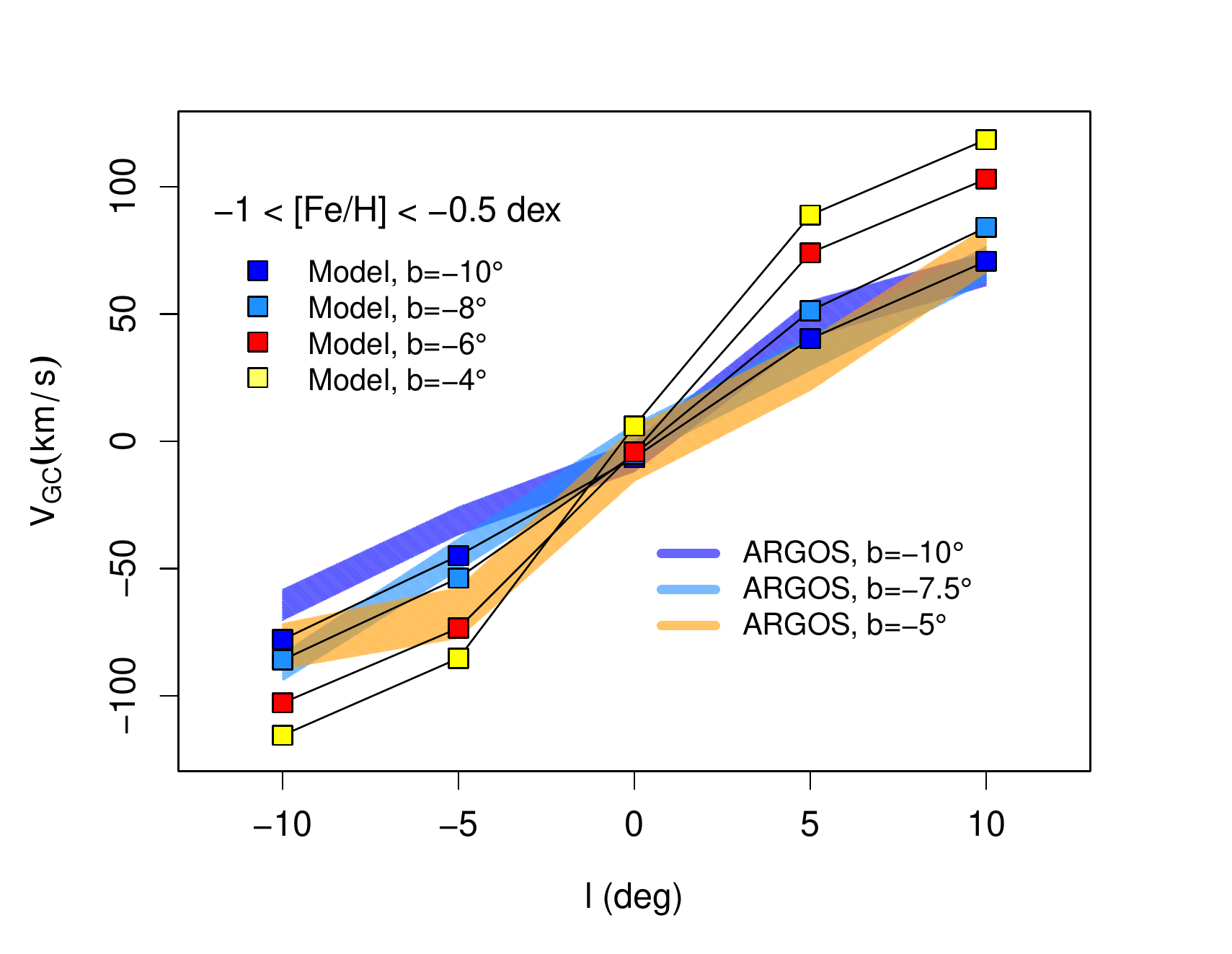}
\includegraphics[clip=true, trim = 30mm 70mm 25mm 75mm, width=1\columnwidth, angle=0]{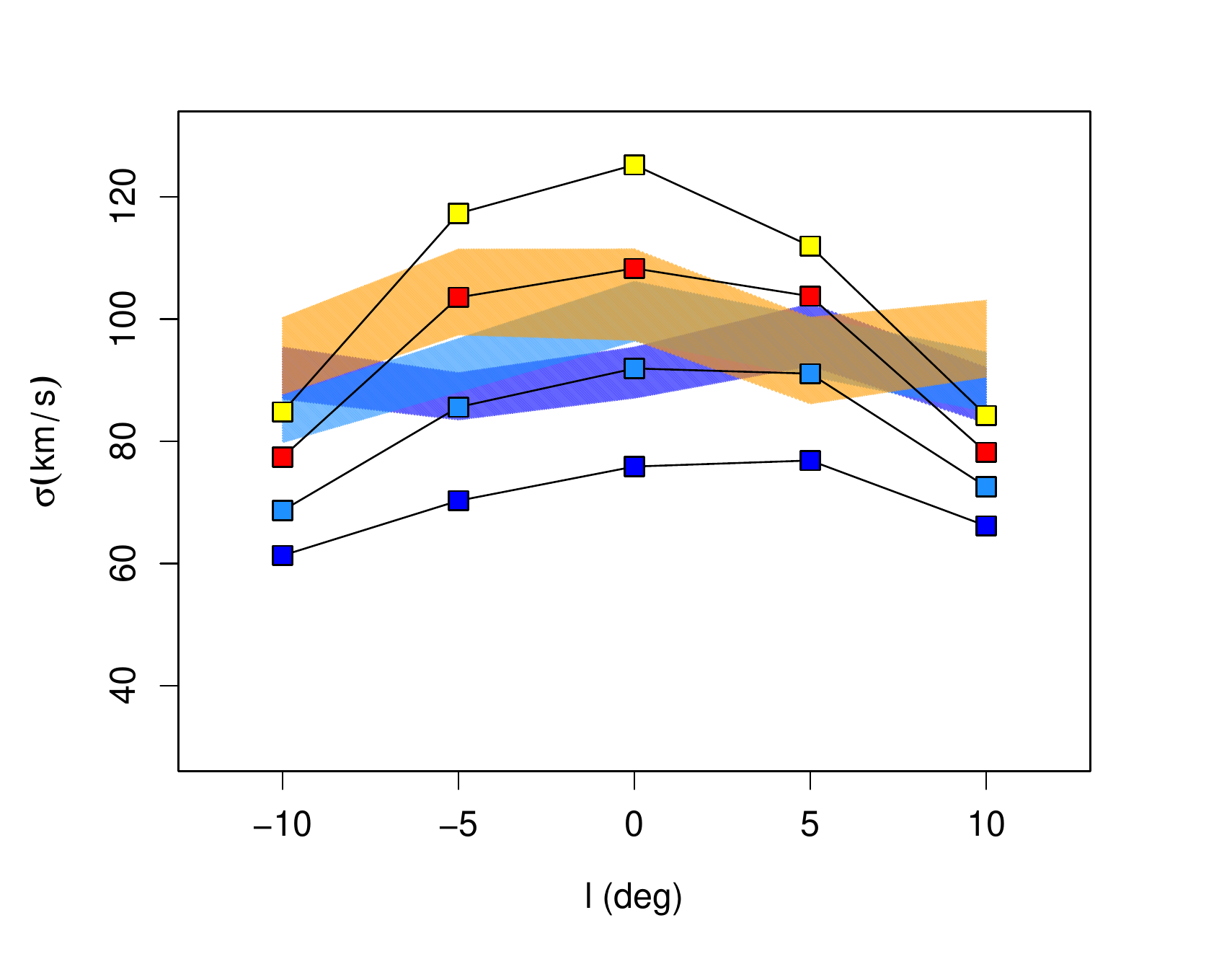}
\caption{From \citet{dimatteo15}: Rotation curves (left panels) and velocity dispersions (right panels) of a boxy bulge formed from a thin stellar disc (square symbols) compared to ARGOS data (colored curves). In the N-body model, only bulge stars are shown. An initial radial metallicity profile [Fe/H] = 0.5?0.4R in the disc is assumed, similar to \citet{martinez13}. Three different metallicity bins are shown, from top to bottom in decreasing [Fe/H], corresponding to the populations A, B, and C, as defined by \citet{ness13spop}. For each plot, four different latitudes are shown for the modeled galaxy: $b = -4^\circ$ (yellow squares), $b = -6^\circ$ (red squares), $b = -8^\circ$ (pale blue squares), $b = -10^\circ$ (dark blue squares). For comparison, ARGOS fields at $b =-5^\circ$ (orange, solid curve), $b = -7.5^\circ$ (pale blue, dashed curve), $b = -10^\circ$ (dark blue, dotted curve) for populations A, B, and C are also given. The thickness of the curves corresponds to the $\pm 1\sigma$ error in the observational data. Note that such model does not reproduce both the cylindrical rotation observed for all the three populations, and the constancy with latitude and longitude of the velocity dispersion of population C.}\label{dimatteo15fig2}
\end{center}
\end{figure*}

\subsubsection*{Failures}
In the scenario where the bulge formed from a pure thin disc, through the intermediate of a bar, it is fundamental that initially -- i.e. before bar formation -- the thin disc has a steep radial metallicity gradient ($\sim -0.4$~dex/kpc as found by \citet{martinez13} and \citet{dimatteo15}). In this way, metal-poor stars originating at large radii in the thin disc are preferentially mapped at high latitudes in the Galactic bulge and give rise to a vertical gradient of about $-0.04$~dex/deg, similar to the findings of \citet{gonzalez13}. In this scenario, because all  bulge stars have a thin disc origin, at a given latitude they should be all part of the boxy/peanut-shaped structure, independently on their metallicities.  This fact has two consequences, which enter in conflict with the detailed structure and chemo-kinematic relations currently known for the bulge components :
\begin{enumerate}
\item at a given latitude, all red clump stars in the Milky Way's bulge with [Fe/H] $> -1$~dex
should show a split in the distribution of their K-magnitudes,
which is not observed; 
\item the metal-poor population (-1~dex$<$ [Fe/H] $\le -0.5$~dex)
should be a kinematically warm replica of the more metal
rich ones (-0.5~dex $<$ [Fe/H]), which is not. 
\end{enumerate}

\begin{figure*}
\begin{center}
\includegraphics[clip=true, trim = 58mm 15mm 48mm 5mm, angle=270,width=1.8\columnwidth, angle=0]{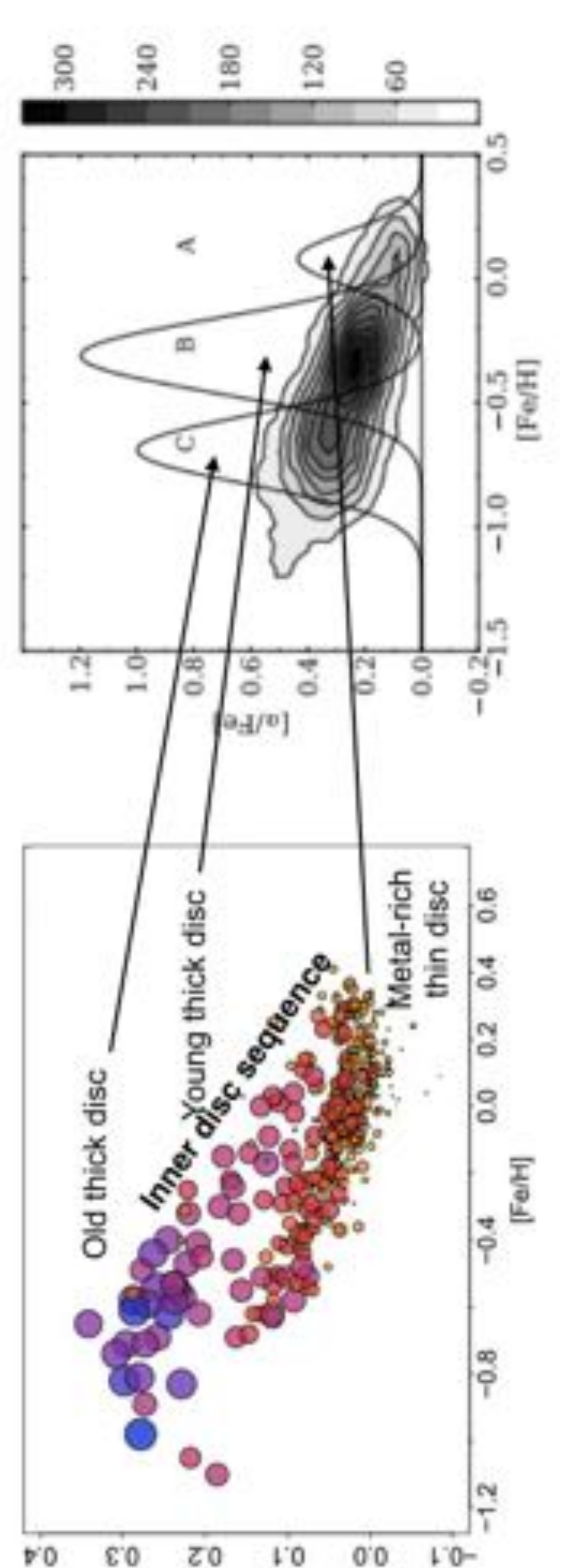}
\caption{The scenario proposed in \citet{dimatteo14, dimatteo15} for the origin of the bulge components A, B and C \citep{ness13spop}. In this scenario, the chemical patterns present in the Galactic bulge can be understood in terms of the chemical patterns of stars at the solar vicinity : the inner (thin+thick) disc sequence is mapped into the boxy bulge, with components A, B and C that correspond, respectively, to the metal-rich thin disc, the young thick disc and the old thick disc, adopting the nomenclature in \citet{haywood13}. Note that in this scenario the outer disc sequence, visible at the solar vicinity, is not present in  the boxy bulge. This latter is indeed representative of the outer disc, not of the inner disc (see Sect.~\ref{thin}), and as a consequence it is not mapped into the boxy bulge. }.\label{bulgescenario}
\end{center}
\end{figure*}

These failures have been presented and discussed in \citet{dimatteo15},  where we have indeed shown how a ``pure thin disc" model can reproduce the global kinematics and chemistry of the bulge, without reproducing the detailed chemo-kinematic and structural properties of its components. The problem is essentially related to the fact that, in such a scenario, metal-poor stars at any given latitude would contribute to the external part of the peanut and that they would have an angular momentum greater than that of more metal-rich stars (whose origin would be in the inner galaxy, given the initial radial metallicity gradient). A higher angular momentum for metal-poor stars would break the cylindrical rotation observed for this component, and would imply also a rotation curve different from that of the most metal-rich populations. This is not observed in the data, which on the contrary show that populations A, B and C all show cylindrical rotation and similar rotation curves. Moreover, because in this scenario metal-poor stars would have a thin disc origin, their velocity dispersion profiles would have a shape similar to those of the most metal-rich components, only shifted at higher absolute values. This again is not observed. 
As shown in Figs.~\ref{dimatteo15fig}, \ref{dimatteo15fig1} and \ref{dimatteo15fig2}, even if successful in explaining global properties of the bulge, the ``pure thin disc" model fails in reproducing the properties of its stellar components (Fig.~\ref{dimatteo15fig2}). For a more extensive discussion on these points, I refer the reader to \citet{dimatteo14} and \citet{dimatteo15}.

\subsection{On the marginal role of any classical bulge in the Milky Way}

If the metal-poor stars in the Milky Way bulge cannot have a thin disc origin, can they belong to a classical bulge ?  Several N-body models, starting from those presented by \citet{shen10}, and then \citet{kunder12, dimatteo14, zoccali14}, show that this is not a viable possibility. They all agree indeed in finding that the mass of any classical bulge present in the Galaxy must be small (10\% at most). \citet{shen10}, for example, showed that for higher mass ratios, the classical bulge would determine signatures in the global velocity dispersion profiles of stars in the bulge not compatible with BRAVA observations \citep[see Fig.~4 in][]{shen10}.  Subsequent works arrived to similar conclusions. Such a small classical bulge can explain neither the large fraction of metal poor-stars observed in the bulge, nor their increasing fraction with latitude. In \citet[][Fig.~11]{dimatteo14}, we have  indeed shown that the contribution of a low-mass classical bulge to the total stellar density is maximal in the innermost regions of the bulge (typically inside 0.5~kpc, where it can account for 20\% at most of the total local stellar density) and this fraction decreases with increasing latitude. The weight of such a classical bulge, and its trends with latitude, are significantly different from those observed for components B and/or C by \citet{ness13spop}, excluding that a low-mass classical bulge can represent the bulk of these populations.\\
Moreover, a classical component would leave signatures in the kinematics of stars in the bulge. Even if stars in the classical bulge can acquire some angular momentum during the bar formation and evolution \citep[see, for example][]{saha12, saha13}, these stars do not attain  rotational velocities comparable to those of disc stars, whatever their location in the boxy structure.  This is shown, for example, in Fig.~12 in \citet{dimatteo14}, but also in \citet{fux97}: in this latter model, the galactocentric radial velocity of the spheroidal component -- whose mass is about 50\% of that of the stellar disc --  is about half the value found for disc stars in all bulge fields (see their Table~5 and their Fig.~16). This rules out  the possibility that the bulk of the metal-poor stars in the Galactic bulge can be associated to a spheroid like that modeled by \citet{fux97}: if this was the case, metal-poor and metal-rich stars should show a significantly different amount of rotation, something excluded by the data \citep{ness13kin, ness15apogee}.

\begin{figure}
\begin{center}
\includegraphics[clip=true, trim = 29mm 20mm 20mm 10mm, width=0.85\columnwidth]{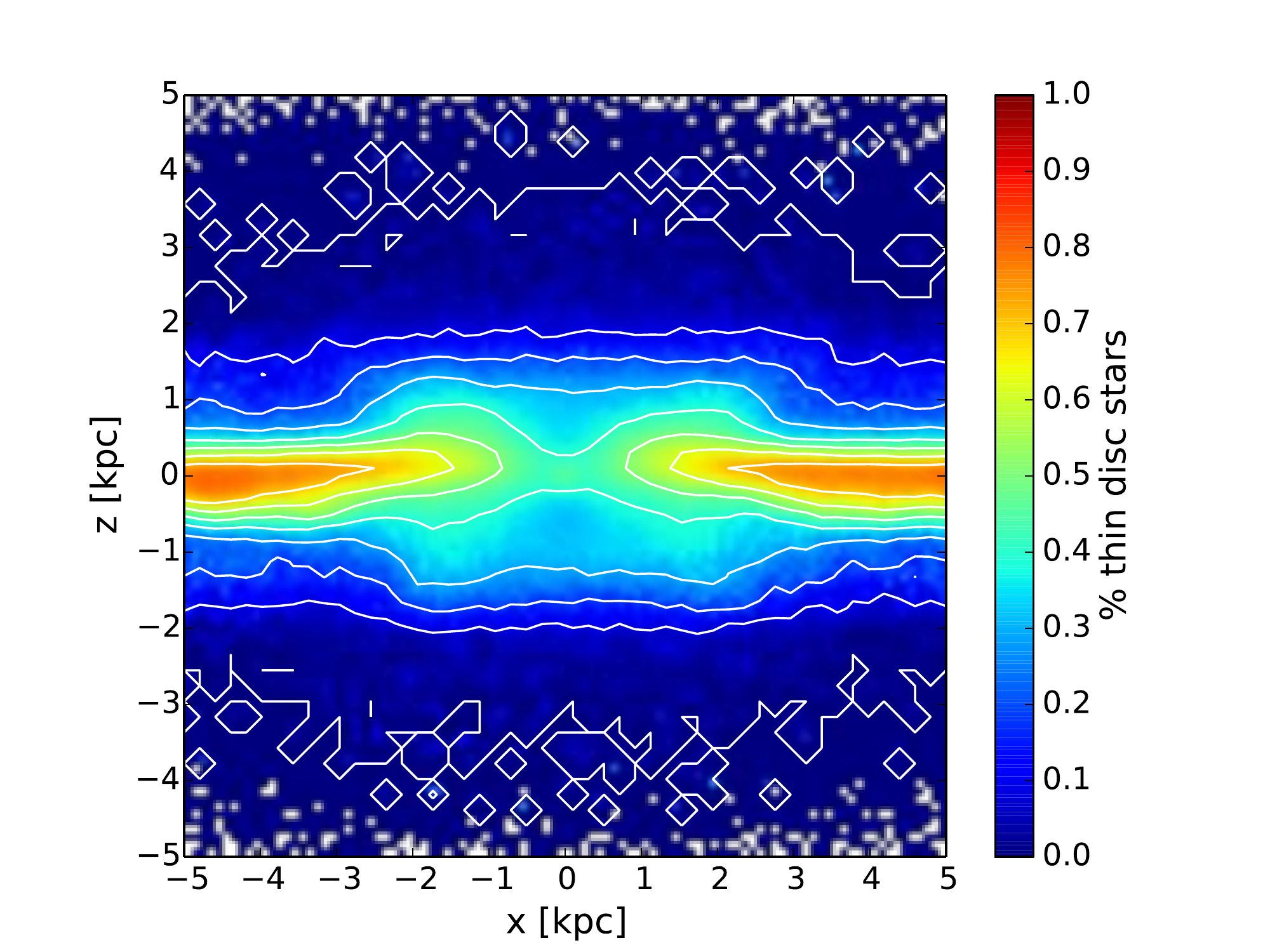}
\includegraphics[clip=true, trim = 29mm 20mm 20mm 10mm, width=0.85\columnwidth]{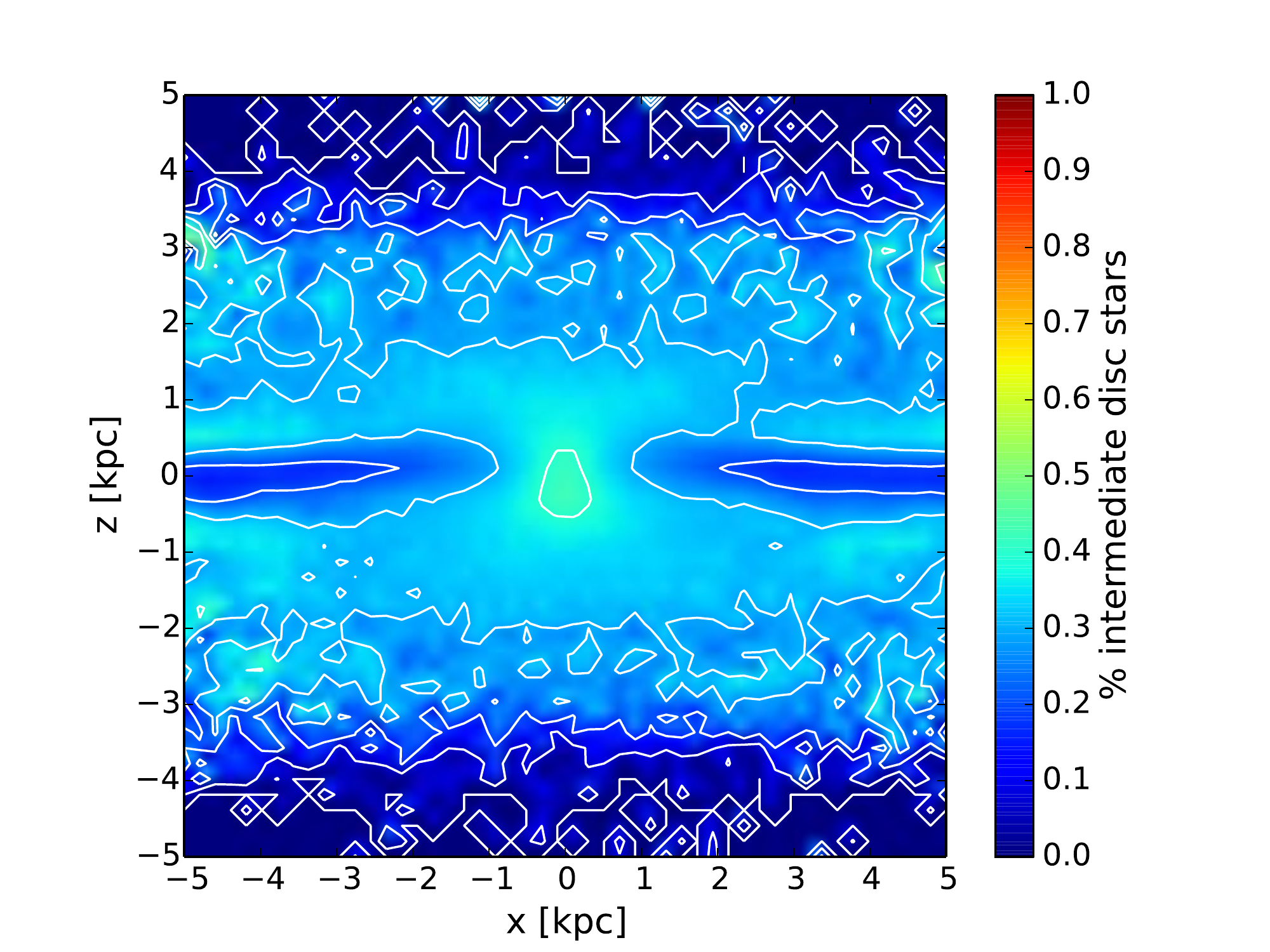}
\includegraphics[clip=true, trim = 29mm 20mm 20mm 10mm, width=0.85\columnwidth]{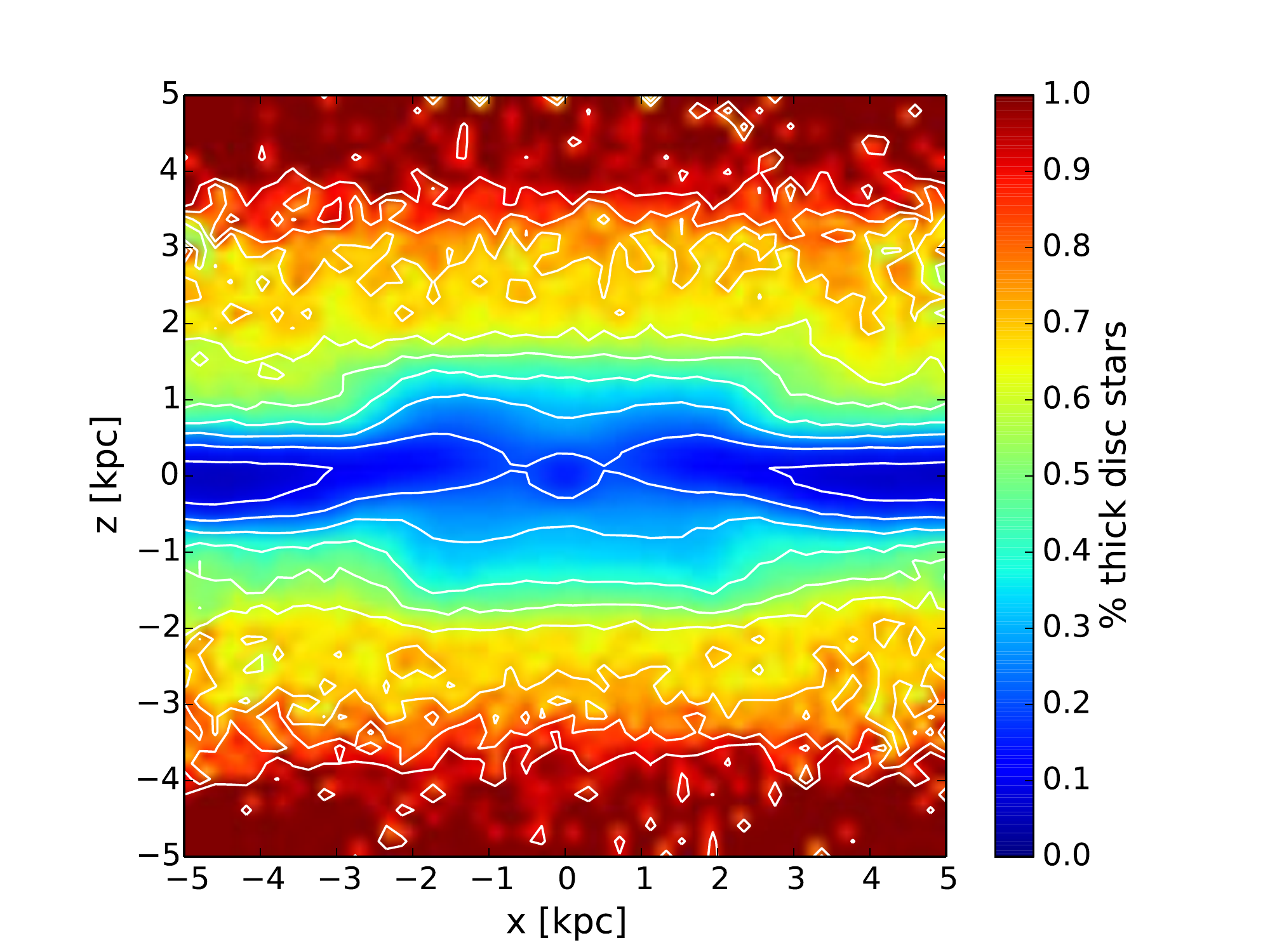}
\caption{The fraction of thin (\emph{top panel}), intermediate (\emph{middle panel}) and thick disc (\emph{bottom panel}) stars in the boxy/peanut-shaped bulge of a simulated Milky Way-type galaxy, seen edge-on.  In this simulation, the three discs components are intended to model the thin disc, young thick disc, and old thick disc, respectively, accordingly to the nomenclature adopted in \citet{haywood13} and summarized here in Fig.~\ref{discscheme}. Note that the fraction of thin disc stars decreases with height above the plane, that of thick disc stars increases with height above the plane, and that of intermediate stars stays nearly constant, with proportions similar to those found for populations A, B and C by \citet{ness13spop}. Each frame is 5~kpc $\times $ 5~kpc in size.}\label{dimatteo16fig2}
\end{center}
\end{figure}

\begin{figure*}
\begin{center}

\includegraphics[clip=true, trim = 0mm 60mm 0mm 50mm, width=2.\columnwidth]{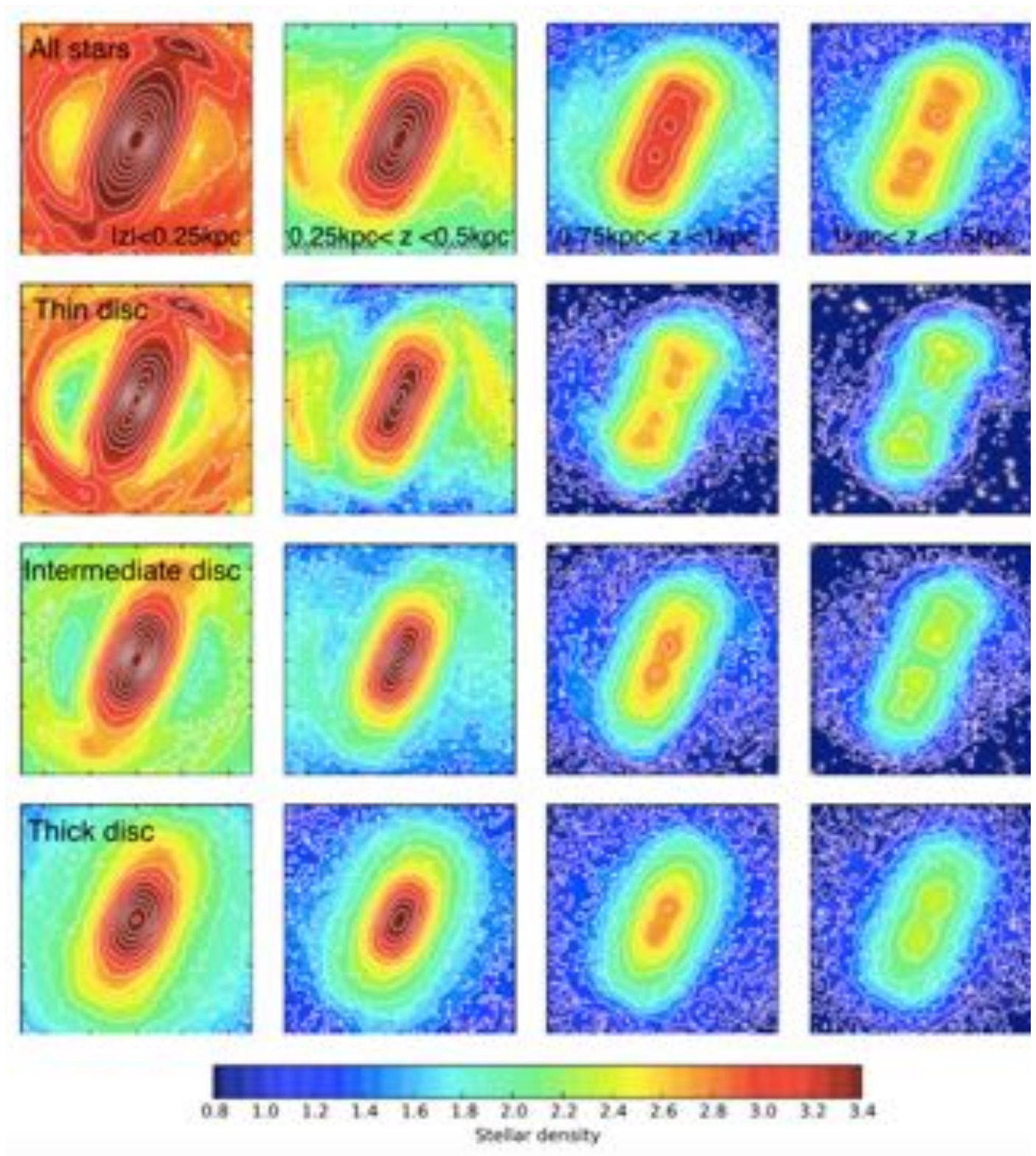}

\caption{Face-on view of the simulated thin+thick disc galaxy for four different slices in $z$. \emph{From left to right} : $|z| < 0.25$~kpc, $0.25~\rm{kpc}<z<0.5~\rm{kpc}$, $0.75~\rm{kpc}<z<1~\rm{kpc}$, $1~\rm{kpc}<z<1.5~\rm{kpc}$. For each of these slices, all stars in the selected $z-$range are shown in the first row. The second, third and last row correspond, respectively, to the distribution of thin, intermediate and thick disc stars only.  I recall that in this simulation the three discs components are intended to model the thin disc, young thick disc, and old thick disc, respectively, accordingly to the nomenclature adopted in \citet{haywood13} and summarized here in Fig.~\ref{discscheme}. Each frame is 5~kpc $\times $ 5~kpc in size.}\label{cutreg2}
\end{center}
\end{figure*}

\subsection{On the bulge/bar/thin+thick disc connection : towards the solution ?}\label{thin+thick}

The previous discussion has shown that neither metal-poor thin disc stars trapped in the bar instability nor a classical bulge can explain the relative weight, fraction as a function of latitude, structure and kinematics of the metal-poor populations observed in the Galactic bulge. At the same time, we have learned from the evidences recalled in Sect.~\ref{stellpop} that a thick disc is massively present in the inner regions of the Milky Way, and that is has [$\alpha$/Fe] and [Fe/H] compatible with those of stars in the bulge.
This  lead us to suggest in \citet{dimatteo14, dimatteo15} that the Milky Way bulge is the result of the mapping of the thin+thick disc of the Galaxy into the boxy/peanut shaped structure. In this scenario, we can understand the chemical patterns present in the Galactic bulge already in terms of the chemical patterns of stars at the solar vicinity : the inner disc sequence is mapped into the boxy bulge, while the outer disc sequence, which -- as we have seen -- contains stars with pericentres slightly inside the solar radius, is not. 
More in details, we suggest that components A, B and C ad defined by \citet{ness13spop} correspond, respectively, to the metal-rich thin disc, the young thick disc and the old thick disc, adopting the nomenclature of \citet{haywood13} and following the scheme presented here in Fig.~\ref{bulgescenario}, and already discussed in \citet{dimatteo14, dimatteo15}. 
Note that this scenario is somehow different from the one presented in \citet{ness13spop}: they indeed associate component C only with the thick disc, while in their interpretation component B should represent the early thin disc. 

How to test the validity of this scheme with N-body simulations ? The recent observational results pose a challenge for idealized models, which have generally represented the stellar content of a Milky Way-type galaxy with a thin stellar disc and an optional bulge. The vertical structuring observed in the data, with a continuous sequence of ``mono-abundance" populations, whose scale height depends on their [$\alpha$/Fe] and [Fe/H], needs to be captured somehow. A solution may be to use, as initial conditions, snapshots of Milky Way-like galaxies taken from cosmological simulations, where ``mono-abundance"  sequences have been indeed reproduced \citep{bird13, stinson13}. The limit of this approach is in the difficulty of investigating a large number of initial conditions, varying thin-to-thick disc ratios, for example, or disc(s) scale lengths and heights. To guarantee to have controlled sequences of (thin+thick) disc models, we have decided to follow a different approach, by constructing idealized galaxies, containing stellar discs of different scale heights and lengths, with masses and sizes in agreement with the most recent estimates for the Milky Way. In the following, I will show how the mapping of a thin+thick disc into a boxy/peanut-shaped bulge is done in one of these simulations, and the kind of trends it predicts. The disc galaxy modeled here is made of 25 million particles, redistributed among stars and dark matter. 
The stellar disc, made of 20 million particles, is structured in three components (hereafter called ``thin", ``'intermediate" and ``thick" discs, respectively) with scale heights varying from 0.3~kpc, to 0.6~kpc and 0.9~kpc, and with velocity dispersions at the solar vicinity similar to those observed for the thin, young thick and old thick disc, respectively \citep{haywood13}. The intermediate and thick disc together constitute 50\% of the stellar mass in the modeled galaxy, with the remaining 50\% being in the thin disc. I would like to highlight here that we chose to model only three disc components simply to identify three phases of an otherwise probably continuous chemical evolution. The same is valid for the Galactic bulge :  components A, B and C have been found by fitting the bulge MDF with multiple gaussians \citep{ness13spop}. But the existence of several components in the bulge MDF does not indicate necessarily that these components are really distinct. And indeed, the [$\alpha$/Fe] versus [Fe/H] plot of the ARGOS stars presented in \citet{ness13spop} shows a continuity between the three components. 

The modeled galaxy is evolved in isolation for 5~Gyr. It rapidly develops a stellar bar which buckles after a couple of Gyrs and forms a boxy/peanut-shaped bulge.
A couple of results of this simulation seem worth mentioning in this context :

\begin{enumerate}
\item Stars of all components contribute to the stellar bar and to the boxy/peanut-shaped structure, but their relative weight depends on the characteristics of the initial stellar disc (i.e. kinematically cold or warm) and on the height above the plane (see Fig.~\ref{dimatteo16fig2}). In particular, in our experiment, by integrating over the whole bulge extent, thin disc stars constitute 40\% of the total stellar densities at $|z|\le$0.25~kpc, but this fraction diminishes to about 25\% at $1.$~kpc$ \le z \le 1.5$~kpc. Thick disc stars show the opposite trend : they represent less than 20\% of the total stellar density at  $|z|\le$0.25~kpc, but their fraction rises to more than 40\% at $1.$~kpc$ \le z \le 1.5$~kpc. Finally, the intermediate disc -- which in our experiment should mimic the young thick disc following \citet{haywood13} nomenclature -- has  a weight which is nearly independent on the vertical distance from the plane : 40\% at $|z|\le$0.25~kpc, 35\% at $1.$~kpc$ \le z \le 1.5$~kpc. Note that at the highest distances from the plane, intermediate plus thick disc stars -- which should mimic the young plus old thick disc, that is components B plus C in our scenario -- constitute the great majority (75\%) of the local stellar density.  Even with the uncertainties still present in the model (thin-to-thick discs mass ratios, and in particular the fraction of young and old thick disc stars to the total thick disc mass; initial velocity dispersion profiles in the inner galaxy, etc), it is striking to see the similarity with the findings of \citet{ness13spop}. In particular the trends found for our thin and thick discs are reminiscent of those of populations A and C in \citet{ness13spop}, while our intermediate disc shows a weight whose nearly independence on the height above the plane is reminiscent of population B. 
\item At a given height above the plane, the boxy/peanut shape is more pronounced in the kinematically cold populations than in the hottest one (see Fig.~\ref{cutreg2}). In particular, in our experiment, a clear peanut-shape in the thick disc is only visible at large heights above the plane, greater than 1~kpc. This has two interesting implications.\\
\citet{ness13spop} found that population C does not show any split in the K-magnitude distribution of stars at latitudes $b=-5^\circ, -7.5^\circ$ and $-10^\circ$, but interestingly at $b=-10^\circ$ the maximum of the K-magnitude distribution of component C does not coincide with the Galaxy centre, but is shifted towards K-magnitude values similar to those where  one of the two maxima of the peanut-shaped populations A and B is found \citep[see Fig.~\ref{ness13fig} here reproduced from][and the previous discussion in Sect.~\ref{bulge}]{ness13spop}. 
This may indicate the presence of a weak peanut-shape also in population C, but possibly visible only at  larger distances from the Galactic plane than those explored by bulge surveys so far. A mapping of the outer bulge, at $|b| > 10^\circ$, would greatly help in understanding the morphology of the metal-poor bulge populations, and to clarify to what extent the characteristics that we have deduced for these  stars  from fields at $|b|\le 10^\circ$ are really representative of their ensemble. \\
The second implication has to do with  the spatial distribution of bulge RR Lyrae stars. \citet{dekany13} found that, at latitudes $b> -5^\circ$, bulge RR Lyrae in the VVV survey do not show a strong barred distribution, as the one traced by metal-rich red clump stars at similar latitudes, but a more spherical shape (note however that, at similar latitudes, \citet{pietrukowicz12} found different results : their sample of RR Lyrae stars from the OGLE-III experiment shows a barred distribution similar to that of bulge red clump giants). It may be tempting to use these findings as indicative of the presence of a classical spheroid in the bulge. However, our model predicts that a different strength in the elongation of the bar in thick and thin disc stars is expected, especially at small heights above the plane. Thus, if a different spatial distribution was found between RR Lyrae with [Fe/H]$\gtrsim -1$~dex --  i.e. the RR Lyrae population with metallicities compatible with the thick disc -- and metal-rich red clump giants in the bulge, this should not be taken as an indication of the presence of a classical spheroid in the bulge. As we show in Fig.~\ref{cutreg2}, indeed, the thicker and kinematically warmer the disc is, the weaker and rounder its stellar bar.  Also the absence of a clear bimodal distribution in RR Lyrae at $b > -5^\circ$ can be understood in a scenario where the metal-rich (i.e. [Fe/H]$\gtrsim -1$~dex) RR Lyrae component is related to the thick disc : indeed, as discussed in the previous point, the peanut-shaped morphology of thick disc stars appears only at large heights above the plane, and is not present at low latitudes, where not even the more metal-rich components of the bulge (A and B)  show any split. 
\end{enumerate}

\section{CONCLUSIONS}

To understand the properties of the Galactic bulge  and the way these properties change with metallicity, we need to place this structure in its environment, that is the Galactic stellar disc, and take into account the most recent findings about the latter. The stellar disc is vertically structured, with an $\alpha-$enhanced, metal-poor and kinematically warm disc (the so-called thick disc) at least as massive as the thin counterpart, and mostly concentrated in the inner regions of the Galaxy. The disc is also radially structured, with chemical patterns that differ in the inner ($R\lesssim 6$~kpc) and outer ($R\gtrsim 10$~kpc) regions. When this radial and vertical structuring are taken into account, the chemical similarity between the bulge and the disc becomes evident, and little room is left for the presence of any classical spheroid. N-body modeling aiming at understanding and explaining the chemo-kinematic and morphological properties of the different bulge components need to implement the new vision we have acquired of the Galactic disc(s) populations. First results about  the modeling of a composite disc galaxy and its mapping into the bulge are presented and support the hypothesis of a pure (thin+thick) disc scenario for the origin of the vast majority (i.e. for stars with [Fe/H]$\gtrsim -1$~dex) of the Milky Way bulge.

\begin{acknowledgements}
These pages are the results of the work done in the last years with many collaborators, and I take this opportunity to thank all of them: F.~Combes, A.~G{\'o}mez, A.~Halle, M.~Haywood, D.~Katz,  M.~Lehnert, B.~Semelin, and O.~Snaith.  In particular, I need  to thank M.~Haywood for introducing me to the universe of the Galactic stellar populations, pushing me to look at the data from a new and fascinating perspective. A.~G{\'o}mez is warmly thanked for the long time spent in producing Milky Way analogs from the simulations. Our science is always the results of many enriching discussions, and the ideas developed in this paper have benefited of the support and criticism of a number of colleagues that I wish to acknowledge : O.~Gerhard, V.~Hill,  I.~Martinez-Valpuesta,  M.Ness, L.~Origlia, A.~Recio-Blanco, M.~Schultheis, M.~Zoccali. A special thank to F.~Matteucci and C.~Morossi, for organizing some of the most enriching meetings I have attended in the last years, and which have deeply contributed to shape the ideas presented in this paper. The ANR (Agence Nationale de la Recherche) is acknowledged for its financial support through the GalHis grant (P.~I. : A.~Robin) till 2013, and now by funding the MOD4Gaia project (ANR-15-CE31-0007). I am grateful to the Editor of this Special Issue on the Galactic bulge, B.~Barbuy, for inviting me to write this contribution.
\end{acknowledgements}


\end{document}